\documentclass[pre,twocolumn,nobalancelastpage,amsmath,amssymb,showpacs,
nofootinbib, superscriptaddress]{revtex4-2}

\usepackage{graphics}      
\usepackage{graphicx}      
\usepackage{longtable}     
\usepackage{url}           
\usepackage{bm}            
\usepackage{amsmath}
\usepackage{dcolumn}
\usepackage{mathtools}
\usepackage{hyperref}
\usepackage[dvipsnames]{xcolor}

\newcommand{\f}{\mathbf{f}}

\newcommand{\bv}{\mathbf{v}}

\newcommand{\br}{\mathbf{r}}

\begin{document}
\title{Comparison of  compression vs  shearing {\color{black}near jamming,} for a simple model of athermal frictionless disks in suspension}
\author{Anton Peshkov}
\affiliation{Department of Physics, California State University Fullerton, Fullerton, CA 92831}
\author{S. Teitel}
\affiliation{Department of Physics and Astronomy, University of Rochester, Rochester, NY 14627}
\date{\today}

\begin{abstract}
Using a simplified model for a non-Brownian suspension, we numerically study  the response of athermal, overdamped, frictionless disks in two dimensions to isotropic and uniaxial compression, as well as to pure {\color{black}and simple} shearing, all at finite constant strain rates $\dot\epsilon$.  We show that isotropic and uniaxial compression result in the same jamming packing fraction $\phi_J$, while  pure shear {\color{black}and simple shear} induced jamming occurs at a slightly higher $\phi_J^*$, consistent with that found previously for simple shearing.  A critical scaling analysis of pure shearing gives critical exponents consistent with those previously found for both isotropic compression and  simple shearing.  
Using orientational order parameters for contact bond directions, we compare the anisotropy of the force and contact networks {\color{black}at both lowest nematic order, as well as higher $2n$-fold order}.  
\end{abstract}
\maketitle


\section{Introduction}

In a recent work \cite{PeshkovTeitel2} we considered isotropic vs uniaxial compression, within a simple granular model of bidisperse non-Brownian spheres in suspension, as a means for numerically studying the effect of stress anisotropy on the jamming transition of frictionless particles.  Isotropic compression at a finite rate results in configurations with an isotropic stress; there is a finite pressure but no shear stress.  Uniaxial compression at a finite rate results in configurations with an anisotropic stress; there is both a finite pressure and a finite shear stress, similar to the case of sheared systems.  Our analysis found that, in three dimensions, jamming via isotropic and uniaxial compression display the same universal critical behavior, despite the difference in stress symmetry.

In this work we consider more generally the differences between   isotropically compressed, uniaxially compressed,  pure sheared, {\color{black}and simple sheared} configurations, when driven at a finite strain rate $\dot\epsilon$, as one approaches and goes above the jamming transition \cite{LiuNagel,OHern,OT1}.  For simplicity we consider the case of circular disks in two dimensions, using the same simple idealized model of a non-Brownian suspension as we used previously \cite{PeshkovTeitel2,PeshkovTeitel}.  

We compare the pressure $p$ and shear stress $\sigma$ arising from such deformations.  Just below the jamming $\phi_J$, we find the pressure from isotropic and uniaxial compression to be  equal within some range of $\phi$ depending on the initial sample preparation.  {\color{black}The pressure from pure shearing and simple shearing are also roughly equal.  However the pressure from pure/simple shearing is roughly an order of magnitude smaller than that from compression.}  The pressures in all {\color{black}four} cases converge as $\phi$ increases above $\phi_J$.   {\color{black}For the shear stress $\sigma$, we again find pure and simple shearing to give the same result.}  Below $\phi_J$, the shear stress for uniaxial compression becomes greater than for pure shear as $\phi$ decreases, while above $\phi_J$ it is reversed.  From a comparison of the stress for these {\color{black}four} deformations, we infer that the jamming $\phi_J$, and the critical exponents at jamming, are the same for isotropic and uniaxial compression.  However we argue that the jamming $\phi_J^*$ for pure {\color{black}and simple} shearing is slightly larger than the $\phi_J$ for compression.  A critical scaling analysis for the case of pure shear{\color{black}, presented in Appendix B,} gives a value for $\phi_J^*$ consistent with that previously found for simple shearing, while the critical exponents are consistent with those found for both simple shearing and for isotropic compression.

We also consider geometrical measures of the configurational contact network, particularly the average number of contacts per particle $Z$, and the fraction of contacts between the different types of particles in our bidisperse system.  We find that, comparing the {\color{black}four types of} deformation, these geometrical measures show small differences  when one is below the jamming $\phi_J$, but that they become equal above $\phi_J$.

Finally, we  compare the system anisotropy that results from uniaxial compression, pure shearing, {\color{black}and simple shearing}.  We show that the stress tensor anisotropy, measured by the macroscopic friction $\mu=\sigma/p$, behaves quite differently {\color{black}when comparing uniaxial compression with pure and simple shearing}.  For {\color{black}pure/simple} shearing $\mu$ is monotonically decreasing as $\phi$ increases, while for uniaxial compression $\mu$ has a sharp minimum at $\phi_J$.  The anisotropy of the contact network, as measured by the fabric tensor, shows similar behavior.  We generalize these anisotropy measures to higher order orientational order parameters of both the force and contact network. 
{\color{black}Comparing uniaxial compression to pure shearing,} we find that the main difference in anisotropy is the relative  magnitude of the isotropic to nematic terms;  higher order orientational moments behave similarly.  {\color{black}In contrast, comparing simple shearing to pure shearing, we find that the isotropic and nematic terms are roughly equal, while the difference is in the higher order moments, which become equal as one approaches and goes above jamming, but become increasingly different as one goes below jamming.}

The remainder of our paper is organized as follows.  In Sec.~\ref{MM} we present our model and numerical methods.  In Sec.~\ref{stress} we present our results for the the system stress, for uniaxial compression, isotropic compression,  pure shearing, {\color{black}and simple shearing}.  In Sec.~\ref{anisotropy} we present a discussion of the anisotropy of the configurational contact and force networks in these cases.  In Sec.~\ref{sum} we summarize our results.  In Appendix A we provide a more complete discussion of the compression ensembles we use, discussing the dependence of the stress on the initial packing fraction $\phi_\mathrm{init}$ from which compression begins, and considering the $\phi_\mathrm{init}\to 0$ limit. 
In Appendix B we provide a more detailed discussion of pure shearing, including a critical scaling analysis.

\section{Model and Methods}
\label{MM}

Our model has been described in detail elsewhere \cite{PeshkovTeitel2,PeshkovTeitel}.  We simulate athermal ($T=0$), bidisperse, frictionless soft-core disks in two dimensions, with equal numbers of big and small disks with diameter ratios $d_b/d_s=1.4$ \cite{OHern}.
Particles, with centers of mass at positions $\br_i$, interact with a one-sided harmonic contact repulsion,
\begin{equation}
U(r_{ij})=\left\{
\begin{array}{cl}
\frac{1}{2}k_e\left(1-\dfrac{r_{ij}}{d_{ij}}\right)^2,&\quad r_{ij}<d_{ij}\\[10pt]
0,&\quad r_{ij}>d_{ij}
\end{array}
\right.
\end{equation}
where $k_e$ is a stiffness constant,  $r_{ij}=|\br_i-\br_j|$ and $d_{ij}=(d_i+d_j)/2$.  The elastic force acting on particle $i$ due to its contact with $j$ is then,
\begin{equation}
\f^\mathrm{el}_{ij}=-\dfrac{dU(r_{ij})}{d\br_i}.
\end{equation}
As a simplified model for particles in solution, we add a dissipative force due to the viscous drag on the particle with respect to the local velocity of the suspending host medium \cite{OT1,OT2,Durian},
\begin{equation}
\f_i^\mathrm{dis} =-k_dV_i\left[\dfrac{d\br_i}{dt}-\bv_\mathrm{host}(\br_i)\right],
\end{equation}
where $k_d$ is a dissipative constant, $V_i$ is the area of particle $i$, and $\bv_\mathrm{host}(\br)$ is the velocity of the host medium at position $\br$.  Particle motion is then determined from these forces using Newton's equation.  We take particle masses to be proportional to their area, $m_i\propto V_i$.  Because our particles are circular and frictionless, we ignore particle rotations.

For the linear  deformations we consider in this work, the background host velocity can be expressed in terms of the strain rate tensor $\dot{\boldsymbol{\epsilon}}$,
\begin{equation}
\mathbf{v}_\mathrm{host}(\mathbf{r}) = \dot{\boldsymbol{\epsilon}}\cdot\mathbf{r}.
\end{equation}
We will consider the {\color{black}particular} cases of uniaxial compression  {\color{black}(uni)}, isotropic compression  {\color{black}(iso)},  pure shear  {\color{black}(ps), and simple shear (ss)}, with respective strain rate tensors,
\begin{align}
&\dot{\boldsymbol{\epsilon}}_\mathrm{uni}=-\left[
\begin{array}{cc}
\dot\epsilon & 0 \\ 0 &0
\end{array}\right],
\quad
\dot{\boldsymbol{\epsilon}}_\mathrm{iso}=-\frac{1}{2}\left[
\begin{array}{cc}
\dot\epsilon & 0 \\ 0 & \dot\epsilon
\end{array}\right],
\label{edef0}
\\[10pt]
&\dot{\boldsymbol{\epsilon}}_\mathrm{ps}=-\frac{1}{2}\left[
\begin{array}{cc}
\dot\epsilon & 0 \\ 0 & -\dot\epsilon
\end{array}\right],
\quad
{\color{black}
\dot{\boldsymbol{\epsilon}}_\mathrm{ss}=\left[
\begin{array}{cc}
0 & \dot\epsilon \\ 0 &0
\end{array}\right]}.
\label{edef}
\end{align}
For uniaxial compression, we compress along the $\mathbf{\hat x}$ direction {\color{black}holding the $\mathbf{\hat y}$ direction fixed}, while for pure shearing we compress along $\mathbf{\hat x}$ while expanding along $\mathbf{\hat y}$ {\color{black}at the same rate}.
Note, the factor of 1/2 in $\dot{\boldsymbol{\epsilon}}_\mathrm{iso}$  is so that the rate of area change is the same for $\dot{\boldsymbol{\epsilon}}_\mathrm{iso}$  as for $\dot{\boldsymbol{\epsilon}}_\mathrm{uni}$.  The factor of 1/2 in $\dot{\boldsymbol{\epsilon}}_\mathrm{ps}$ is so that we can then view uniaxial compression as a superposition of an isotropic compression plus a pure shear,
\begin{equation}
\dot{\boldsymbol{\epsilon}}_\mathrm{uni}=\dot{\boldsymbol{\epsilon}}_\mathrm{iso}+\dot{\boldsymbol{\epsilon}}_\mathrm{ps}.
\label{esupo}
\end{equation}
{\color{black}The simple shear can be viewed as a pure shear of rate $\dot\epsilon$ with compression along the $(1,-1)$ diagonal, combined with a rotation of the system with angular velocity $-(\dot\epsilon/2)\mathbf{\hat z}$ \cite{MHT}.}  

\begin{figure}
\centering
\includegraphics[width=2.5in]{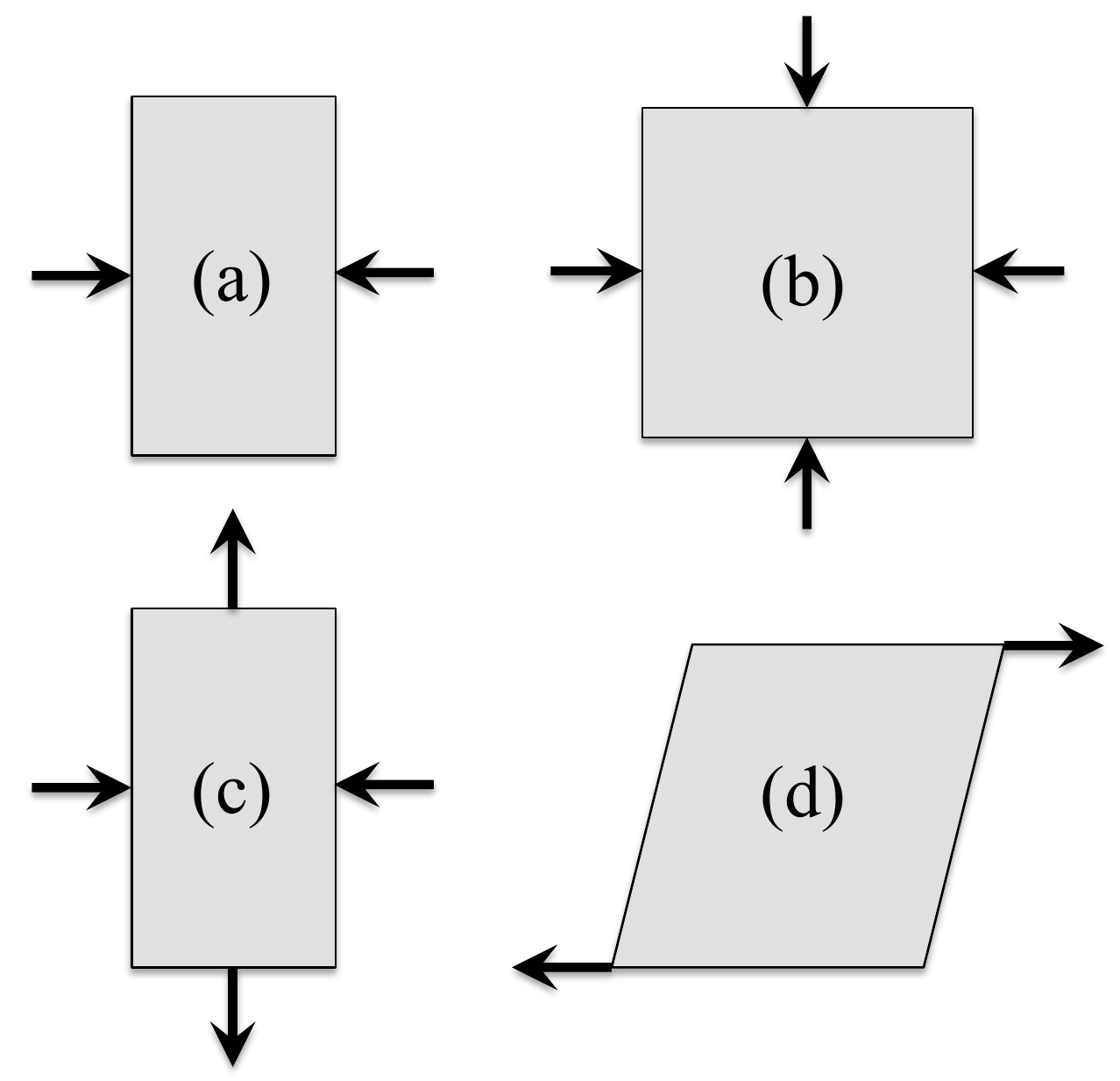}
\caption{{\color{black}The geometry of the four linear deformations described by the strain rate tensors of Eqs.~(\ref{edef0}) and (\ref{edef}): (a) uniaxial compression, (b) isotropic compression, (c) pure shear, and (d) simple shear.}
}
\label{strains}
\end{figure}

Our particles are placed in a rectangular box with side lengths $\mathbf{L}=(L_x,L_y)$, centered at $\mathbf{r}=0$.  
{\color{black}For uniaxial compression, isotropic compression, and pure shear, as}
 we make our elastic deformations, the box lengths vary according to
\begin{equation}
\dfrac{d\mathbf{L}}{dt}=\dot{\boldsymbol{\epsilon}}\cdot\mathbf{L}.
\end{equation}
At each integration step, particles that would fall outside the system box are returned to the box using periodic boundary conditions \cite{PeshkovTeitel2}.  {\color{black}For simple shear, the box lenghts $(L_x,L_y)$ stay constant, and the box skews to a rhomboidal shape at a fixed rate, with Lees-Edwards boundary conditions being applied \cite{Lees}.  A sketch showing the geometry of our four linear deformations is shown in Fig.~\ref{strains}.}

To carry out our numerical simulations, we recast our model in terms of three dimensionless parameters \cite{Vag}.  The first is the packing fraction $\phi$,
\begin{equation}
\phi = \dfrac{1}{L_xL_y}\sum_i V_i.
\end{equation}
For both isotropic and uniaxial compression, $\phi$ increases with time as the system gets compressed.  For the area-preserving pure {\color{black} and simple} shear deformations, $\phi$ stays constant.

The second is the quality factor $Q$, which measures the strength of the dissipative force relative to the elastic force.  If $\tau_d=m_s/k_dV_s$ and $\tau_e=\sqrt{m_sd_s^2/k_e}$ are the time scales associated with the dissipative and elastic forces \cite{Vag}, we have,
\begin{equation}
Q=\dfrac{\tau_d}{\tau_e}=\dfrac{\sqrt{m_sk_e}}{k_dV_sd_s}
\end{equation}
As $Q$ decreases, inertial effects decrease.   For $Q$ sufficiently small, behavior becomes independent of the particular value of $Q$ and one enters the overdamped limit corresponding to massless particles, $m_s\to 0$ \cite{Vag,VagbergOlssonTeitel}.  For our simulations we will use $Q=1$, which is sufficiently small to put us in this overdamped limit \cite{Vag}.   

In the $m_s\to 0$ overdamped limit both $\tau_e$ and $\tau_d\to 0$, however we can  define a time scale that remains finite \cite{Vag},
\begin{equation}
\tau_0  =\dfrac{\tau_e^2}{\tau_d}=\dfrac{\tau_e}{Q}= \,\dfrac{k_dV_sd_s^2}{k_e}.
\end{equation}
Our third dimensionless parameter is then the dimensionless strain rate,
\begin{equation}
\dot\epsilon\tau_0.
\end{equation}
Henceforth we will take our unit of length to be $d_s=1$, and our unit of time to be $\tau_0=1$.  Quoted values of $\dot\epsilon$ are therefore the same as $\dot\epsilon\tau_0$.  We consider strain rates spanning the range $\dot\epsilon=10^{-8.5}$ to $10^{-4}$.

We use LAMMPS \cite{lammps} to integrate the equations of motion, using a time step of $\Delta t/\tau_0=0.1$.  
Unless otherwise noted, we use $N=32768$ total particles.  
Our simulations for isotropic and uniaxial compression start with an initial configuration at low packing $\phi_\mathrm{init}=0.4$, constructed as follows.  We place particles down one by one at random, but making sure that there are no particle overlaps; if an overlap occurs, we discard that particle and try again until all $N$ particles are placed in the box.  
For  isotropic compression, we use a square box with $L_x=L_y$.  For uniaxial compression we start at $\phi_\mathrm{init}$ with a rectangular box with $L_y < L_x$, such that the box becomes roughly square by the time we have compressed to the jamming $\phi_J$.  For  uniaxial and isotropic compression we average our results over 20 independent initial configurations.
For our pure shear simulations we start at each $\phi$ with a configuration generated by our uniaxial compression protocol, using the same strain rate $\dot\epsilon$.
For each $\dot\epsilon$ {\color{black}we shear to a total strain $\epsilon\sim 1$, dropping an initial $\epsilon\sim 0.2$ to reach the sheared steady-state, and then averaging over the remainder of the run (see Appendix B for details).}
We then average our results over 10 independent initial configurations, except for our slowest rates $\dot\epsilon\le 10^{-7}$, where we use only a single initial configuration.
{\color{black}For simple shear, we start at each $\phi$ with a totally random initial configuration, and energy relax the system without shearing for a time $10^6\Delta t$ to remove any initial unphysically large particle overlaps.  We then shear the system, discarding the first $ \epsilon\sim1$ to reach the sheared steady-state, and then average over an additional strain of $\epsilon\sim9$.}

\section{Results: Stress}
\label{stress}

In this section we consider the stress generated in the system by the elastic deformations.  We consider only the  stress  arising from the elastic forces, since this is the dominant term at low strain rates.  The stress tensor can be expressed in terms of the force moments as \cite{OHern},
\begin{equation}
\mathbf{P}=\dfrac{1}{L_xL_y}\left\langle \sum_{i<j}\f_{ij}^\mathrm{el}\otimes(\br_i-\br_j)\right\rangle,
\label{estress}
\end{equation}
where $\langle\cdots\rangle$ denotes an average over our independent runs {\color{black}for compression, and an average over both strain and independent runs for shearing.}

A dimensionless stress tensor can be defined as \cite{Vag},
\begin{equation}
\mathbf{p}=\dfrac{\tau_e^2}{m_s}\,\mathbf{P}.
\end{equation}
{\color{black}The stress tensor may be written in the general form,
\begin{equation}
\mathbf{p}=\left[
\begin{array}{cc}
p+\delta p & p_{xy} \\ p_{xy} & p-\delta p
\end{array}
\right],
\label{eptensor}
\end{equation}
}where the pressure $p$ is the isotropic part of the stress, {\color{black}   given by the average of the eigenvalues of $\mathbf{p}$.  The anisotropic part of the stress is given by the deviatoric stress $\sigma$, determined as half the difference of the eigenvalues,
\begin{equation}
\sigma=\sqrt{\delta p^2+p_{xy}^2}
\label{devstress}
\end{equation}
We will refer to $\sigma$ as simply the shear stress.}

For isotropic compression, symmetry gives {\color{black}that the stress tensor is isotropic, and so} $\sigma = 0$.  For  both uniaxial compression and pure shear, {\color{black}as in Fig.~\ref{strains}, symmetry requires the stress tensor be diagonal, so that $p_{xy}=0$ and the shear stress is $\sigma=\delta p$.  For simple shearing, if our system were a uniform elastic continuum one would find $\delta p=0$ and so $\sigma=|p_{xy}|$.  For our granular system we will find that, while $p_{xy}\gg\delta p$, $\delta p$ does not strictly vanish, and so $\sigma$ is given by the full expression of Eq.~(\ref{devstress}).

The pressure $p$ and shear stress $\sigma$ thus give two parameters characterizing the stress tensor.  The third parameter needed to completely specify the stress tensor can be taken to be the orientation of the maximal stress axis, given by the eigenvector of the maximal eigenvalue of $\mathbf{p}$.  We denote this by $\vartheta_2$, the angle of the maximal stress axis with respect to the $\mathbf{\hat x}$ direction.  For our uniaxial compression and pure shear, where the maximal stress direction is along $\mathbf{\hat x}$ (see Fig.~\ref{strains}(a)(c)), symmetry requires $\vartheta_2=0$.  For simple shear, since $p_{xy}\gg\delta p$, we have $\vartheta_2\approx -\pi/4$.
}

The area-preserving process of  shearing at a finite rate defines a steady-state ensemble of configurations that becomes independent of the initial starting configuration, provided one shears sufficiently long.  This has previously been observed for the case of simple shearing \cite{Vagberg.PRE.2011}, and  in Appendix B we  confirm that it is also the case for pure shearing.  Our results below for $p$ and $\sigma$ for {\color{black}both pure and simple} shearing represent a time average over configurations, once this steady-state limit has been reached.  The resulting values of $p$, $\sigma$ {\color{black} and $\vartheta_2$} are determined solely by the parameters $\phi$, $Q$ and $\dot\epsilon$.  

For isotropic and uniaxial compression, however, the situation is not as simple.  As one compresses, $\phi$ increases, and the ensemble of configurations one passes through can depend on the ensemble of initial configurations one starts the compression from.  In our case, where we start from configurations of non-overlapping particles at an initial packing $\phi_\mathrm{init}$, our values of $p$ and $\sigma$  can depend not only on the parameters $\phi$, $Q$, and $\dot{\boldsymbol{\epsilon}}$, but also on the additional parameter $\phi_\mathrm{init}$.  In Appendix A we consider this dependence of the stress on $\phi_\mathrm{init}$.  We find that as $\phi_\mathrm{init}$ decreases, the stress for both isotropic and uniaxial compression approaches a well defined $\phi_\mathrm{init}\to 0$ limit.  For $\phi_\mathrm{init}$ not too small, the resulting $p(\phi)$ and $\sigma(\phi)$ approach this limiting curve as $\phi$ reaches the dense limit, just below jamming.  Since using very small $\phi_\mathrm{init}$ can be  computationally expensive for the large $N=32768$ system size that we wish to use, to avoid finite size effects near jamming, here we use $\phi_\mathrm{init}=0.4$.  We find that this $\phi_\mathrm{init}$ is sufficiently small that our results are roughly independent of $\phi_\mathrm{init}$ once $\phi \gtrsim 0.80$.  Further details are presented in Appendix A.

In Fig.~\ref{p-s-vs-phi}(a) we plot $p$ vs $\phi$, for several different strain rates $\dot\epsilon$, for our four types of deformation:  uniaxial compression, isotropic compression,  pure shear, {\color{black}and simple shear}. 
The vertical dashed line in this figure shows the jamming transition for isotropic compression, $\phi_J=0.8415$, as we have determined previously \cite{PeshkovTeitel}.
As  found before \cite{PeshkovTeitel2,OT1,PeshkovTeitel,OT2}, we find here (not shown) that {\color{black}all four} deformations have a linear rheology, $p\propto\dot\epsilon$, provided one is below and not too close to jamming.  Above jamming, $p$ approaches a constant as $\dot\epsilon\to 0$.  Comparing the {\color{black}four} cases, 
for $\phi \lesssim \phi_J$, we see that $p_\mathrm{iso}$ is equal to $p_\mathrm{uni}$, {\color{black}$p_\mathrm{ps}$ is equal to $p_\mathrm{ss}$, but the shearing pressure} is about a factor 10 smaller than {\color{black}that of compression}.  For $\phi\gtrsim\phi_J$, however, we see that the $p$ for all {\color{black}four} cases are becoming equal as $\dot\epsilon\to 0$.

In Fig.~~\ref{p-s-vs-phi}(b) we similarly plot $\sigma$ vs $\phi$ at different $\dot\epsilon$, for uniaxial compression, pure shear, {\color{black}and simple shear} ($\sigma_\mathrm{iso}=0$ by symmetry).  As with the pressure $p$, the shear stress $\sigma\propto\dot\epsilon$ if one is not too close to jamming, while $\sigma$ approaches a constant as $\dot\epsilon\to 0$ above jamming.  Here we see that that, {\color{black}as with the pressure, $\sigma_\mathrm{ps}$ is equal to $\sigma_\mathrm{ss}$.  Comparing shearing to uniaxial compression, we see that}    $\sigma_\mathrm{uni}$ and $\sigma_\mathrm{ps}$ are generally of the same order of magnitude, but $\sigma_\mathrm{ps} < \sigma_\mathrm{uni}$ for $\phi \lesssim\phi_J$, while $\sigma_\mathrm{ps} > \sigma_\mathrm{uni}$ for $\phi \gtrsim\phi_J$.  As $\dot\epsilon \to 0$, $\sigma_\mathrm{ps}/\sigma_\mathrm{uni} \to 1$ at $\phi\approx\phi_J$.  

\begin{figure}
\centering
\includegraphics[width=3.4in]{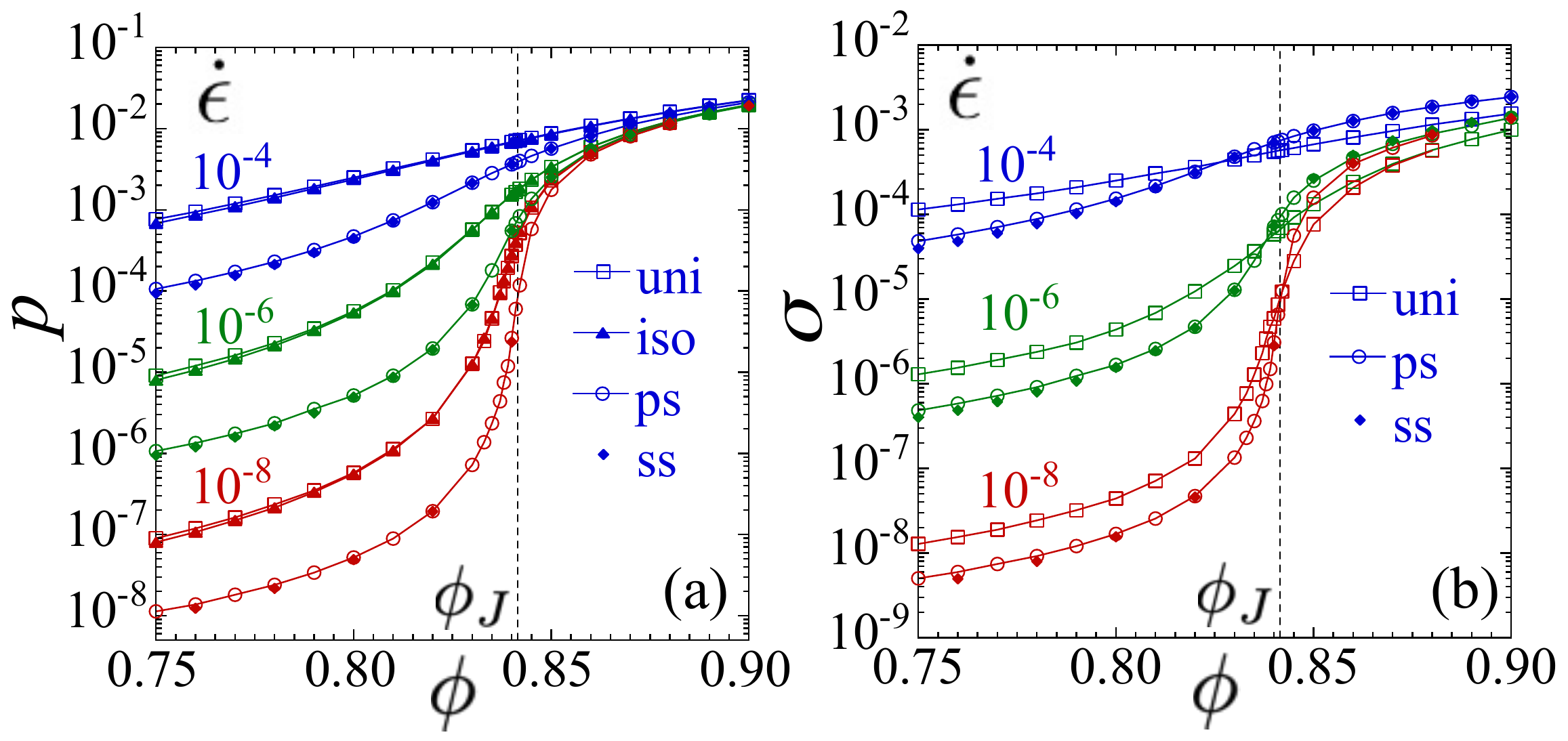}
\caption{(a) Pressure $p$, and (b) shear stress $\sigma$, vs packing $\phi$ for uniaxial compression $(\square)$, isotropic compression $(\blacktriangle)$,  pure shear $(\circ)$, and {\color{black}simple shear $(\blacklozenge)$} of $N=32768$ particles, at different strain rates $\dot\epsilon=10^{-4}$, $10^{-6}$ and $10^{-8}$. 
The vertical dashed line indicates the isotropic compression-driven jamming $\phi_J = 0.8415$.  The estimated error in the data is typically smaller than the size of each data point symbol. For the shear stress in (b) $\sigma_\mathrm{iso}=0$ and is not shown.
}
\label{p-s-vs-phi}
\end{figure}

\begin{figure}
\centering
\includegraphics[width=3.4in]{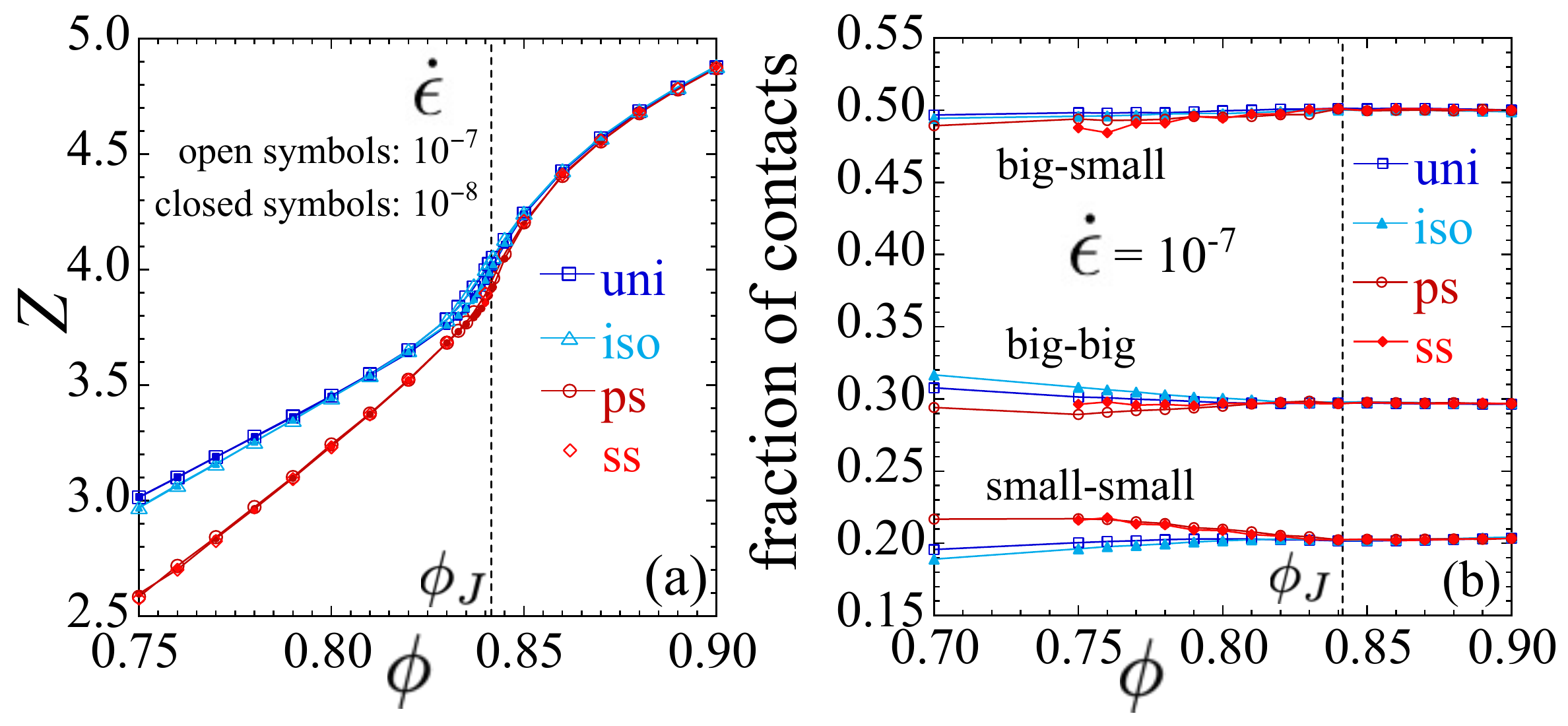}
\caption{(a) Average number of contacts per particle $Z$ vs $\phi$ for uniaxial compression, isotropic compression,  pure shear, {\color{black}and simple shear}  at the two strain rates $\dot\epsilon=10^{-7}$ and $10^{-8}$.  Rattler particles are included in the calculation of $Z$.  (b) Fraction of the contacts that are between two small particles, two big particles, and a big and small particle, vs $\phi$ for uniaxial compression, isotropic compression,  pure shear, {\color{black}and simple shear} at $\dot\epsilon=10^{-7}$. The vertical dashed line indicates the isotropic compression-driven jamming $\phi_J = 0.8415$.  The system has $N=32768$ particles.  The estimated error in the data is typically smaller than the size of each data point symbol.
}
\label{Z-contacts-vs-phi}
\end{figure}

Next we consider some geometrical properties of our configurations. In Fig.~\ref{Z-contacts-vs-phi}(a) we plot the average number of contacts per particle $Z$ vs $\phi$ for our  {\color{black}four} types of deformation, at the two small strain rates $\dot\epsilon=10^{-7}$ and $10^{-8}$.   As was observed before for isotropic compression \cite{PeshkovTeitel2} and simple shearing \cite{Heussinger1,OlssonRelax,Olsson3D}, we find that, as $\dot\epsilon\to 0$,  $Z$ stays finite and varies roughly linearly with $\phi$ for $\phi<\phi_J$, while at $\phi_J$ and above we see the square root singularity, $(Z-Z_J)\sim (\phi-\phi_J)^{1/2}$ associated with jamming \cite{PeshkovTeitel2,OHern,Wyart}.  Here our value of $Z_J$ at jamming is slightly below the isostatic value of $Z_\mathrm{isostatic}=2d=4$  since, for simplicity, we have not excluded  rattler particles when computing $Z$ \cite{LiuNagel,OHern}.  

Our observation that $Z$ approaches a constant as $\dot\epsilon\to 0$ indicates that, at low strain rates, the system forms a well defined contact network at all packings, even below $\phi_J$.  The extent of the particle overlaps at these contacts varies $\propto\dot\epsilon$, giving rise to the linear rheology in $p$ below $\phi_J$, however the geometry of the contact network remains the same.  This is the hard-core limit.
Similar to $p$, the contact numbers $Z$ for the {\color{black}four} cases appear to become equal for $\phi\gtrsim\phi_J$, but differ below $\phi_J$, where $Z_\mathrm{iso}$ is roughly equal to $Z_\mathrm{uni}$, while $Z_\mathrm{ps}{\color{black}=Z_\mathrm{ss}}$ is  noticeably smaller than the other two.

In Fig.~\ref{Z-contacts-vs-phi}(b) we plot the fraction of the particle contacts that are between two small particles, two big particles, and between one small and one big particle vs $\phi$ for the strain rate $\dot\epsilon=10^{-7}$.  Similar to $Z$ and $p$, we see that above $\phi_J$ these fractions become equal for all  cases.  Below $\phi_J$ we see that pure {\color{black}and simple} shearing produce more small-small contacts and fewer big-big contacts than does isotropic or uniaxial compression.  Our above results thus show that, in the jammed state above $\phi_J$, it is only the shear stress $\sigma$ that clearly distinguishes between {\color{black}compressive vs shearing} deformations.

\begin{figure}
\centering
\includegraphics[width=3.4in]{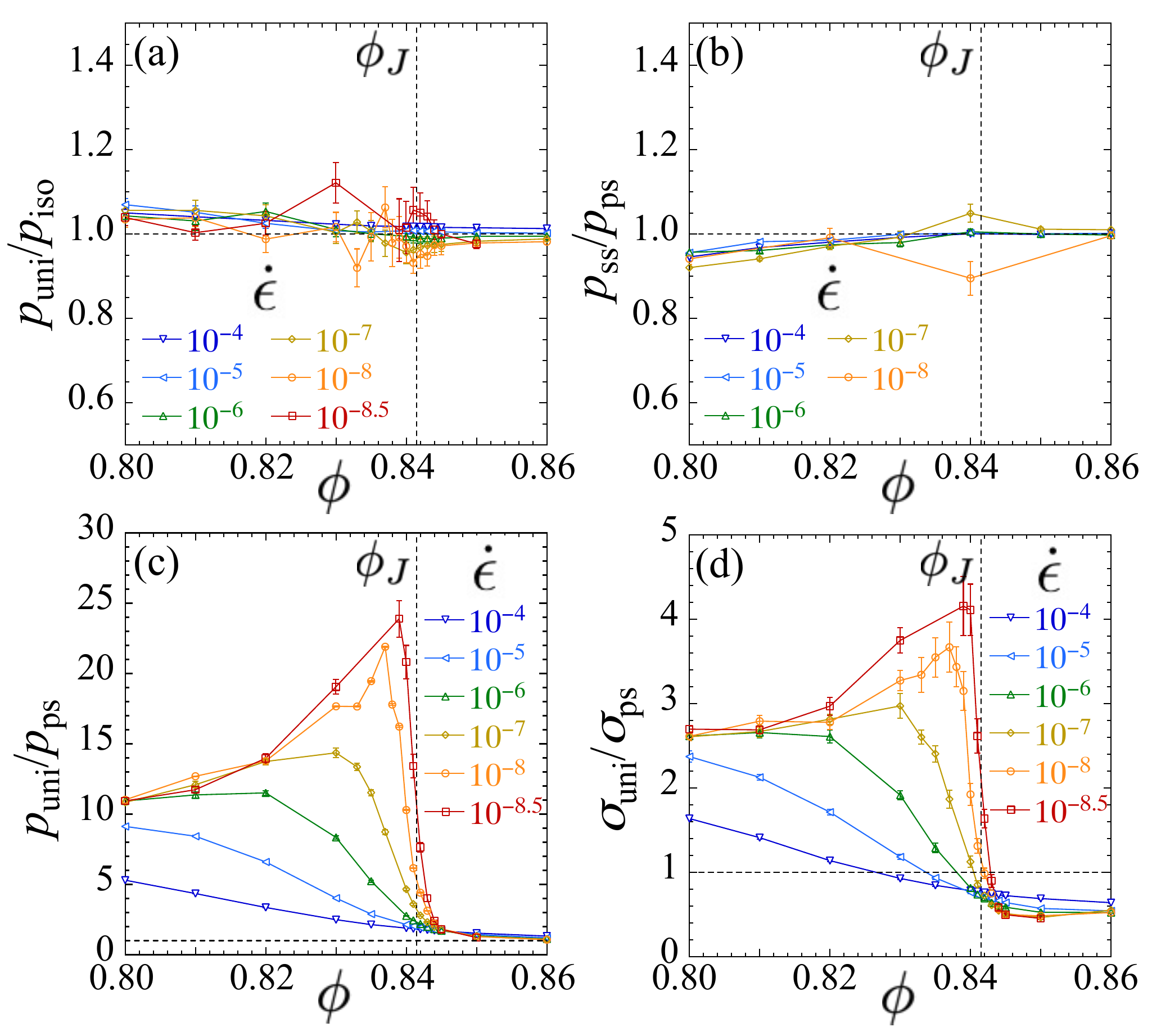}
\caption{Stress ratios comparing uniaxial compression, isotropic compression,  pure shear, {\color{black} and simple shear}. Pressure ratios (a) $p_\mathrm{uni}/p_\mathrm{iso}$, (b) {\color{black}$p_\mathrm{ss}/p_\mathrm{ps}$}, (c) $p_\mathrm{uni}/p_\mathrm{ps}$, and (d) shear stress ratio $\sigma_\mathrm{uni}/\sigma_\mathrm{ps}$ vs $\phi$ for different strain rates $\dot\epsilon$.  The vertical dashed line indicates the isotropic compression-driven jamming $\phi_J = 0.8415$. The system has $N=32768$ particles.
}
\label{Rp-Rs-vs-phi}
\end{figure}

Next we consider whether all {\color{black}four} cases  jam at exactly the same critical packing fraction $\phi_J$.  To investigate this we compute the stress ratios between the different cases.  In Fig.~\ref{Rp-Rs-vs-phi}(a) we {\color{black}compare uniaxial with isotropic compression, plotting} $p_\mathrm{uni}/p_\mathrm{iso}$ vs $\phi$, for several different strain rates $\dot\epsilon$.  
We see no particular features as $\phi$ passes through $\phi_J$.  Since, as $\dot\epsilon\to 0$, the bulk viscosity $\zeta\equiv p/\dot\epsilon\sim (\phi_J-\phi)^{-\beta}$ diverges at $\phi_J$ with the  critical exponent $\beta$, the absence of any features in $p_\mathrm{uni}/p_\mathrm{ps}$ near $\phi_J$ strongly suggests that $p_\mathrm{uni}$ and $p_\mathrm{iso}$ jam at exactly the same $\phi_J$ and their $\zeta$ diverge with the same exponent $\beta$.
This conclusion is in agreement with what we explicitly demonstrated for three dimensions in an earlier work \cite{PeshkovTeitel2}.  

In Fig.~\ref{Rp-Rs-vs-phi}(b) we   {\color{black} compare pure with simple shearing, plotting $p_\mathrm{ss}/p_\mathrm{ps}$ vs $\phi$.  We similarly see no particular features as $\phi$ passes through $\phi_J$.  The same behavior is found if we look at $\sigma_\mathrm{ss}/\sigma_\mathrm{ps}$.  This suggests that pure and simple shearing jam at the same packing, with the same critical exponent $\beta$. }

In Fig.~\ref{Rp-Rs-vs-phi}(c) we compare {\color{black}compression with shearing, plotting $p_\mathrm{uni}/p_\mathrm{ps}$ vs $\phi$.}  Here we see a very different behavior.
We find that $p_\mathrm{uni}/p_\mathrm{ps}$ develops a peak just below {\color{black}the isotropic compression-driven} $\phi_J$; as $\dot\epsilon$ decreases, the height of this peak increases and the location of the peak moves closer to $\phi_J$.  The same behavior is found for  $\sigma_\mathrm{uni}/\sigma_\mathrm{ps}$ in ~\ref{Rp-Rs-vs-phi}(d).

Two possible explanations for such behavior are: (i) {\color{black}Shearing} jams at a slightly higher $\phi_J^*$ than the $\phi_J$ for compression.  In this case we would expect, in the limit $\dot\epsilon\to 0$, that $p_\mathrm{uni}/p_\mathrm{ps}$ diverges as $\phi\to\phi_J$ from below, stays infinite for $\phi_J<\phi<\phi_J^*$, and approaches a finite constant for all $\phi>\phi_J^*$. (ii)  {\color{black}Shearing} jams at the same $\phi_J$ as does compression, but with a smaller exponent $\beta^*<\beta$.  In this case we would expect, in the limit $\dot\epsilon\to 0$, that $p_\mathrm{uni}/p_\mathrm{ps}$ diverges as $\phi\to\phi_J$ from below, but approaches a finite constant for all $\phi>\phi_J$.  Our data is more consistent with the possibility (i), since we see that there remains a small interval above the compressive $\phi_J$ where the stress ratio continues to increase as $\dot\epsilon$ decreases.  Prior work \cite{OT2,OT3} has demonstrated that for simple shearing  our model jams at the packing $\phi_J^*=0.8435>\phi_J=0.8415$, and that the exponent $\beta$ is the same as found for compression \cite{PeshkovTeitel}.  In Appendix B we present a detailed critical scaling analysis of our pure shearing data that confirms that pure shearing indeed behaves the same as simple shearing, with {\color{black}the same} $\phi_J^*>\phi_J$, {\color{black}and}  the same exponent $\beta$.

We have previously noted in Eq.~(\ref{esupo}) that, with regard to the strain rate tensor $\dot{\boldsymbol{\epsilon}}$, uniaxial compression can be regarded as a superposition of isotropic compression plus pure shearing.  It is therefore natural to wonder whether a similar superposition holds for the resulting stresses in these flowing states, if one is in the region where the rheology is linear.  Our results in Fig.~\ref{Rp-Rs-vs-phi}, however, show that it does not.  Were a superposition of stress to hold, we would expect $p_\mathrm{uni}=p_\mathrm{iso}+p_\mathrm{ps}$ and $\sigma_\mathrm{uni}=\sigma_\mathrm{ps}$ (since $\sigma_\mathrm{iso}=0$).  Our results in Fig.~\ref{Rp-Rs-vs-phi}, as well as our earlier results in Ref.~\cite{PeshkovTeitel} (see Fig.~1(b) of that work), show that for $\dot\epsilon\le 10^{-7}$ we remain in the linear rheology region for $\phi$ up to at least $\sim 0.82$.  However below $\phi=0.82$ we see from Figs.~\ref{Rp-Rs-vs-phi}(a) and \ref{Rp-Rs-vs-phi}(d) that $p_\mathrm{uni}\approx p_\mathrm{iso}$  and $\sigma_\mathrm{uni}\approx 2.5\sigma_\mathrm{ps}$.  In general, we see from Fig.~\ref{Rp-Rs-vs-phi}(d)  that $\sigma_\mathrm{uni}=\sigma_\mathrm{ps}$ only at an isolated point close to $\phi_J$.  Thus there is no principle of superposition for stress in the flowing states below $\phi_J$.

\section{Results: Anistropy}
\label{anisotropy}

In this section we {\color{black} consider the three elastic deformations that result in stress anisotropic systems, uniaxial compression, pure shearing, and simple shearing, and compare different measures of that anisotropy.}

{\color{black}\subsection{Stress and Fabric Anisotropy}}

\begin{figure}
\centering
\includegraphics[width=3.4in]{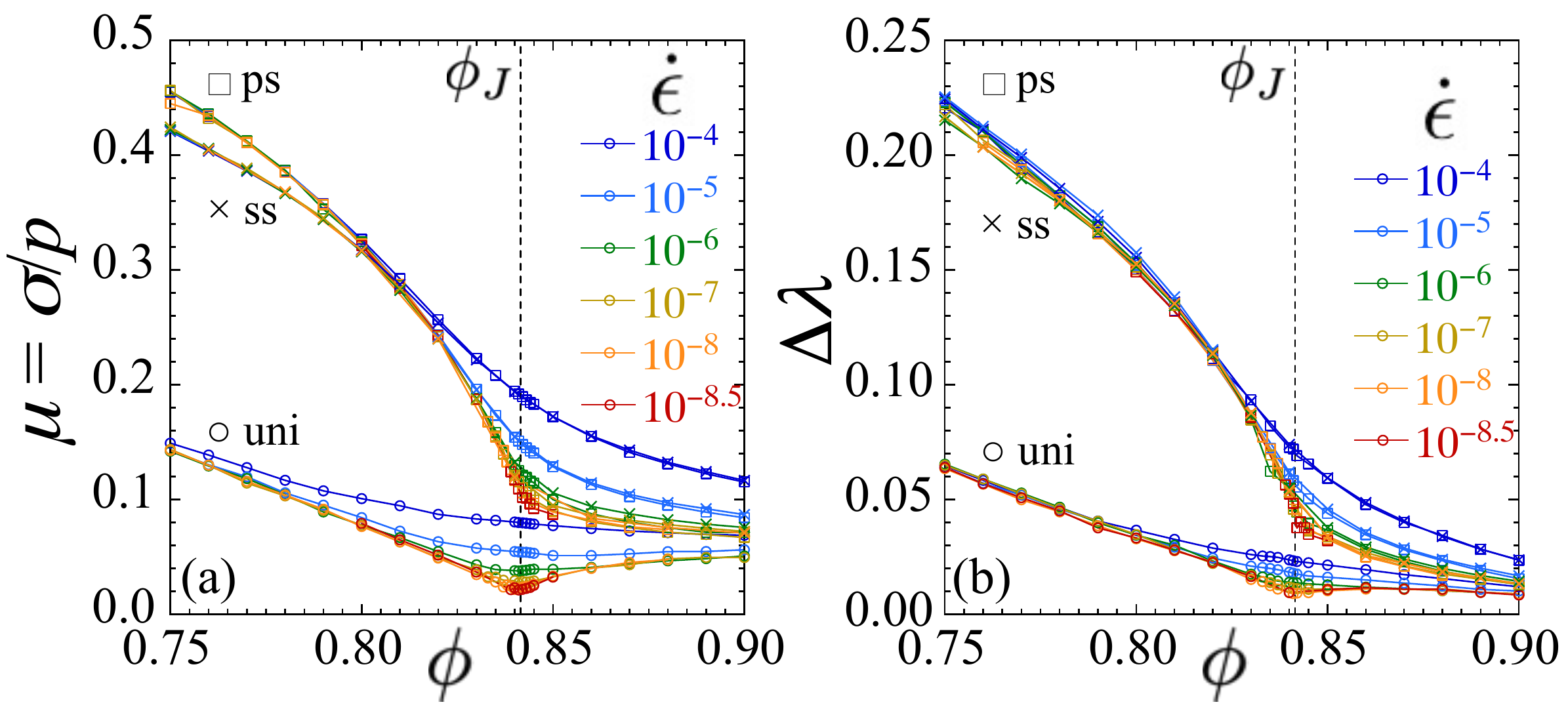}
\caption{(a) Stress tensor anisotropy $\mu=\sigma/p$, and (b) fabric tensor anisotropy $\Delta\lambda$ vs $\phi$ for different strain rates $\dot\epsilon$, {\color{black}for uniaxial compression ($\circ$), pure shearing ($\square$) and simple shearing ($\times$). The strain rates vary from $\dot\epsilon=10^{-4}$ to $10^{-8.5}$ as the curves go from top to bottom; for simple shear the slowest rate is $10^{-8}$.}
The vertical dashed line indicates the isotropic compression-driven jamming $\phi_J = 0.8415$. The system has $N=32768$ particles.  The estimated error in the data is typically smaller than the size of each data point symbol.
}
\label{mu-lambda-vs-phi}
\end{figure}

We first consider the  anisotropy of the stress tensor, parameterized by the macroscopic friction $\mu\equiv \sigma/p$.  In Fig.~\ref{mu-lambda-vs-phi}(a) we plot $\mu$ vs $\phi$, for different strain rates $\dot\epsilon$, {\color{black}for these three cases}. 
In all cases $\mu$ approaches a limiting, finite valued, curve as $\dot\epsilon\to 0$.  
{\color{black}We see that both pure and simple shearing give equal results, $\mu_\mathrm{ps}=\mu_\mathrm{ss}$, except at the smaller $\phi$ below jamming, where $\mu_\mathrm{ps}$ is slightly larger.}

However, as we noted previously for three dimensions \cite{PeshkovTeitel2},  we find a distinct difference {\color{black}comparing uniaxial compression with shearing.}
For  {\color{black}both pure and simple} shearing the $\dot\epsilon\to 0$ limiting curve is monotonically decreasing as $\phi$ increases,  as was seen previously for simple shearing \cite{Vagberg.PRL.2014}. In contrast, for uniaxial compression, this curve develops a cusp-like minimum at $\phi_J$.  For  shearing we find the value of $\mu$ at $\phi_J$ to be, $\mu^J_\mathrm{shear}\approx 0.1$, while  for uniaxial compression we find $\mu^J_\mathrm{uni}\approx 0.02$.  Thus the ratio $\mu^J_\mathrm{shear}/\mu^J_\mathrm{uni}\approx 5$.  However, the key point is that in {\color{black}all} cases $\mu^J$ stays finite; the system remains anisotropic at jamming and above.

For $\phi<\phi_J$, the smaller value of $\mu_\mathrm{uni}$ for uniaxial compression is primarily due to the much larger pressure {\color{black}present in uniaxial compression compared to shearing}.  As seen in Fig.~\ref{Rp-Rs-vs-phi}(c), close below $\phi_J$ we have $p_\mathrm{uni}/p_\mathrm{shear}\approx 10$.  In contrast, as seen in Fig.~\ref{Rp-Rs-vs-phi}(d), the shear stress is $\sigma_\mathrm{uni}/\sigma_\mathrm{shear}\approx 2.7$. So $\mu_\mathrm{shear}/\mu_\mathrm{uni}\approx 3.7$; upon approaching $\phi_J$, this ratio increases.  Above $\phi_J$, we see from Fig.~\ref{Rp-Rs-vs-phi}(c) that $p_\mathrm{uni}/p_\mathrm{shear}\approx 1$, while $\sigma_\mathrm{uni}/\sigma_\mathrm{shear}\approx 0.5$.  Thus, above $\phi_J$ we have $\mu_\mathrm{shear}/\mu_\mathrm{uni}\approx 2$, and this difference is now due entirely to the difference in the shear stress.

{\color{black}Finally we can ask about the direction $\vartheta_2$  of the maximal stress axis.  As noted earlier, for uniaxial compression and pure shear, symmetry requires $\vartheta_2=0$; the maximal stress axis is located along the compressive $\mathbf{\hat x}$ direction.  For simple shear we find (details below, see Fig.~\ref{theta-thetah-vs-phi-1234}) that $\vartheta\approx -\pi/4$.}

It is interesting to ask how much of the difference in anisotropy, comparing uniaxial compression to  shearing, is due to anisotropy in the  force network, as measured by the stress tensor, vs how much is due  to the geometrical anisotropy of the contact network.
We therefore consider the behavior of the fabric tensor \cite{fabric}.  If $\mathbf{\hat r}_{ij}=(\mathbf{r}_i-\mathbf{r}_j)/r_{ij}$ is the unit vector pointing along a bond connecting two particles in contact, and $M_{b}=NZ/2$ is the total number of contact bonds in the configuration, the fabric tensor can be defined as,
\begin{equation}
\boldsymbol{\mathcal{F}}=\left\langle\dfrac{1}{M_{b}}\sum_{(i,j)}\mathbf{\hat r}_{ij}\otimes\mathbf{\hat r}_{ij}\right\rangle,
\end{equation}
where the sum is over all bonds $(i,j)$ in the contact network.
Since, for circular particles, the elastic contact force $\mathbf{f}_{ij}^\mathrm{el}=f_{ij}^\mathrm{el}\,\mathbf{\hat r}_{ij}$ is always parallel to $\mathbf{\hat r}_{ij}$, if we define the force-moment as $h_{ij}\equiv f_{ij}^\mathrm{el}\,r_{ij}$, we can rewrite the stress tensor of Eq.~(\ref{estress}) as,
\begin{equation}
\mathbf{P}=\left\langle\dfrac{1}{L_xL_y} \sum_{(i,j)}h_{ij}\,\mathbf{\hat r}_{ij}\otimes\mathbf{\hat r}_{ij}\right\rangle .
\label{ep2}
\end{equation}
We thus see that the fabric tensor is similar to the stress tensor, but without weighting each bond by its force-moment.  The fabric tensor is thus   a purely geometric measure  of the contact network.

If we define $\theta_{ij}$ as the angle $\mathbf{\hat r}_{ij}$ makes with respect to 
$\mathbf{\hat x}$, then we can write,
\begin{align}
\boldsymbol{\mathcal{F}}&=\left[
\begin{array}{cc}
\langle \cos^2\theta_{ij}\rangle & \langle\cos\theta_{ij}\sin\theta_{ij}\rangle \\[10pt]
\langle\cos\theta_{ij}\sin\theta_{ij}\rangle & \langle\sin^2\theta_{ij}\rangle
\end{array}
\right]
\\[12pt]
&=\dfrac{1}{2}\mathbf{I}+\dfrac{1}{2}\left[
\begin{array}{cc}
\langle \cos2\theta_{ij}\rangle & \langle\sin2\theta_{ij}\rangle \\[10pt]
\langle \sin2\theta_{ij}\rangle & -\langle\cos2\theta_{ij}\rangle
\end{array}
\right]
\label{eF2}
\end{align}
where $\mathbf{I}$ is the identity tensor, and now $\langle\cdots\rangle$ represents a combined average over both bonds within a given configuration and over different independent configurations.  
The first piece $\mathbf{I}/2$ is the isotropic part of $\boldsymbol{\mathcal{F}}$, while the second piece gives the anisotropic part.

The eigenvalues of $\boldsymbol{\mathcal{F}}$ are then,
\begin{equation}
\lambda_\pm = \dfrac{1}{2}\left(1\pm\Delta\lambda\right),
\end{equation}
where
\begin{equation}
\Delta\lambda=\dfrac{\lambda_+-\lambda_-}{\lambda_++\lambda_-}=\sqrt{\langle\cos 2\theta_{ij}\rangle^2 +\langle\sin 2\theta_{ij}\rangle^2}.
\label{eDlambda}
\end{equation}
{\color{black}The quantity $(\lambda_++\lambda_-)/2=1$ is the analog of the pressure $p$, while $(\lambda_+-\lambda_-)/2$ is the analog of the shear stress $\sigma$.  Thus} we see that $\Delta\lambda$ for the fabric tensor is analogous to $\mu=\sigma/p$ for the stress tensor.  {\color{black}The angle of the maximal eigenvector with respect to $\mathbf{\hat x}$ we will denote by $\theta_2$.}

For uniaxial compression and pure shearing, the reflection symmetry of the deformations $y\leftrightarrow -y$,  implies the symmetry $\theta_{ij}\leftrightarrow -\theta_{ij}$.  This leads to the conclusion that $\langle \sin 2\theta_{ij}\rangle=0$.   Thus $\Delta\lambda = |\langle\cos 2\theta_{ij}\rangle|$ {\color{black}and $\theta_2=0$.  For simple shearing, there is no such symmetry and one must use the full expression of Eq.~(\ref{eDlambda}).  Similar to the stress tensor, we  find that for simple shearing $\theta_2\approx -\pi/4$.}

In Fig.~\ref{mu-lambda-vs-phi}(b) we plot $\Delta\lambda$ vs $\phi$ for different strain rates $\dot\epsilon$, for  uniaxial compression,  pure shearing, {\color{black}and simple shearing}.  We see qualitatively the same behavior as found for $\mu$.  
We find {\color{black} $\Delta\lambda_\mathrm{ps}=\Delta\lambda_\mathrm{ss}$, with some small deviations at the lower $\phi$.  Both}
$\Delta\lambda_\mathrm{ps}$ {\color{black}and $\Delta\lambda_\mathrm{ss}$ are} monotonically decreasing as $\phi$ increases, while $\Delta\lambda_\mathrm{uni}$ has a cusp-like minimum at $\phi_J$, and $\Delta\lambda^J_\mathrm{ps}/\Delta\lambda^J_\mathrm{uni}\approx 5$.  The close correspondence of the behavior of $\Delta\lambda$ with that of $\mu$ suggests that the geometry of the contact network  is the primary mechanism for the anisotropy in the systems.

{\color{black}\subsection{Orientational Order Parameters}

The fabric anisotropy $\Delta\lambda$ of Eq.~(\ref{eDlambda})  can also be viewed as the magnitude of the nematic order parameter for contact  bond directions \cite{Donev0}.  Here we generalize to higher order moments of the anisotropy, by considering the full angular distribution of bond forces and directions.  Henceforth, we will refer to the set of contact bond  directions $\{\mathbf{\hat r}_{ij}\}$ as the ``contact network."  We will refer to the set of force-moments $\{h_{ij}\mathbf{\hat r}_{ij}\}$ as the ``force network."

Let $\mathcal{P}(\theta,h)$ be the joint probability distribution that a given contact bond is in direction $\theta$ and has a force-moment $h$.
Because of the symmetry, $\mathbf{\hat r}_{ij}=-\mathbf{\hat r}_{ji}$ and $h_{ij}=h_{ji}$, this distribution has the periodicity $\mathcal{P}(\theta+\pi,h)=\mathcal{P}(\theta,h)$.  We therefore define $\mathcal{P}(\theta,h)$ as a function on the range $\theta\in[-\pi/2,\pi/2)$ only, and normalize it appropriately.  We can then define,
\begin{equation}
\mathcal{P}(\theta)=\int_0^\infty\!\!\!dh\,\mathcal{P}(\theta,h)
\end{equation}
as the probability density to have a contact bond at angle $\theta$, independent of its force-moment $h$, and,
\begin{equation}
\tilde h(\theta)=\int_0^\infty\!\!\!dh\,\mathcal{P}(\theta,h)h
\end{equation}
as the average force moment per radian at angle $\theta$.

The function $\tilde h(\theta)$  incorporates in its definition the probability that there will indeed be a contact bond at angle $\theta$.  Alternatively we can ask, what is the average force-moment on a bond at angle $\theta$, independent of the likelihood that there is a bond at that orientation.  Writing the joint distribution as $\mathcal{P}(\theta,h)=\mathcal{P}(h|\theta)\mathcal{P}(\theta)$, where $\mathcal{P}(h|\theta)$ is the conditional probability to find a force-moment $h$, given that there is a bond at $\theta$, we can then define,
\begin{equation}
h(\theta) = \int_0^\infty\!\!\!dh\,\mathcal{P}(h|\theta)h = \dfrac{\tilde h(\theta)}{\mathcal{P}(\theta)}.
\end{equation}
To illustrate the difference between $h(\theta)$ and $\tilde h(\theta)$, image that all bonds had the same force-moment $h$;
then we would have $\tilde h(\theta)= h \mathcal{P}(\theta)$, while $h(\theta)=h$ would be constant.

We can then expand $\mathcal{P}(\theta)$ in terms of a Fourier series \cite{Azema}.  We have,
\begin{align}
\mathcal{P}(\theta)&=\dfrac{1}{\pi}+\dfrac{2}{\pi}\sum_{n=1}\big[A_n\cos 2n\theta+B_n\sin 2n\theta\big]\\
&=\dfrac{1}{\pi}+\dfrac{2}{\pi}\sum_{n=1}S_{2n}\cos(2n[\theta-\theta_{2n}])
\label{eFT}
\end{align}
where the Fourier coefficients are given by,
\begin{align}
A_n&=\int_{-\pi/2}^{\pi/2}\!\!\!\!d\theta\,\mathcal{P}(\theta)\cos 2n\theta =\langle\cos 2n\theta\rangle\\[10pt]
B_n&=\int_{-\pi/2}^{\pi/2}\!\!\!\!d\theta\,\mathcal{P}(\theta)\sin 2n\theta =\langle\sin 2n\theta\rangle\\[10pt]
S_{2n}&=\sqrt{A_n^2+B_n^2}=\sqrt{\langle\cos2n\theta\rangle^2+\langle\sin2n\theta\rangle^2}
\label{eS2n}
\end{align}
and $\theta_{2n}$ is given by,
\begin{equation}
\tan(2n\theta_{2n})=\dfrac{B_n}{A_n}=\dfrac{\langle\sin2n\theta\rangle}{\langle\cos2n\theta\rangle}.
\label{eth2n}
\end{equation}
The magnitude and orientation $(S_{2n},\theta_{2n})$ is just the $2n$-fold  orientational order parameter for the bond directions of the geometrical contact network \cite{Donev0}.  Odd order orientational order parameters (i.e., $S_{2n+1}$) all vanish due to the symmetry $\mathcal{P}(\theta)=\mathcal{P}(\theta+\pi)$.  

Comparing with Eq.~(\ref{eDlambda}), we see that  $S_2\equiv\Delta\lambda$; the fabric tensor  anisotropy $\Delta\lambda$ is the same as the magnitude of the nematic order parameter $S_2$ of the contact network. The higher moment $S_4$ gives the tetratic order,  while $S_6$ gives the hexatic order, etc.  Considering the distribution $\mathcal{P}(\theta)$, and its Fourier coefficients $S_{2n}$, thus generalizes the fabric tensor to higher order orientational moments.

We can similarly expand $\tilde h(\theta)$ in a Fourier series, to get,
\begin{align}
\tilde h(\theta)&=\dfrac{C_0}{\pi}+\dfrac{2}{\pi}\sum_{n=1}\big[C_n \cos2n\theta+D_n\sin2n\theta\big]
\label{ehFT0}
\\[10pt]
&=C_0\left[\dfrac{1}{\pi}+\dfrac{2}{\pi}\sum_{n=1}\mathbb{S}_{2n}\cos(2n[\theta-\vartheta_{2n}])\right],
\label{ehFT}
\end{align}
where the Fourier coefficients are given by,
\begin{align}
C_n&=\int_{-\pi/2}^{\pi/2}\!\! d\theta\,\tilde h(\theta)\cos 2n\theta=\langle h\cos2n\theta\rangle
\\[12pt]
D_n&=\int_{-\pi/2}^{\pi}\!\! d\theta\,\tilde h(\theta)\sin2n\theta=\langle h\sin2n\theta\rangle.
\end{align}
Note, $C_0=\langle h\rangle$ is just the average force-moment.
The magnitude $\mathbb{S}_{2n}$ is then given by,
\begin{equation}
\mathbb{S}_{2n}=\dfrac{\sqrt{C_n^2+D_n^2}}{C_0}=\dfrac{\sqrt{\langle h\cos2n\theta\rangle^2+\langle h\sin2n\theta\rangle^2}}{\langle h\rangle},
\label{eSp2n}
\end{equation}
and the orientation $\vartheta_{2n}$ is given by,
\begin{equation}
\tan(2n\vartheta_{2n})=\dfrac{D_n}{C_n}=\dfrac{\langle h\sin2n\theta\rangle}{\langle h\cos2n\theta\rangle}.
\label{ethp2n}
\end{equation}
The magnitude and orientation $(\mathbb{S}_{2n},\vartheta_{2n})$  therefore gives the $2n$-fold orientational order parameter for the force network.

We can now relate the above to the stress tensor.  Using the definition of $\mathbf{p}$ in Eq.~(\ref{ep2}), and making the corresponding steps that led to Eq.~(\ref{eF2}), we can write,
\begin{equation}
\mathbf{p}=\dfrac{M_b}{2L_xL_y}\left[
\begin{array}{cc}
\langle h\rangle+\langle h\cos 2\theta\rangle & \langle h\sin 2\theta\rangle \\[12pt]
 \langle h\sin 2\theta\rangle &\langle h\rangle -\langle h\cos 2\theta\rangle
\end{array}
\right].
\end{equation}
Comparing with Eq.~(\ref{eptensor}) we then have,
\begin{equation}
\begin{array}{c}
p=\frac{M_b}{2L_xL_y}\langle h\rangle,\quad \delta p = \frac{M_b}{2L_xL_y}\langle h\cos 2\theta\rangle, \\[10pt]
p_{xy}=\frac{M_b}{2L_xL_y}\langle h\sin 2\theta\rangle,
\end{array}
\end{equation}
and so, from Eq.~(\ref{devstress}), we get,
\begin{equation}
\mu=\dfrac{\sigma}{p}=\dfrac{\sqrt{\langle h\cos 2\theta\rangle^2+\langle h\sin 2\theta\rangle^2}}{\langle h\rangle} = \mathbb{S}_2.
\end{equation}
Thus $\mathbb{S}_2\equiv \mu$ is  the nematic order parameter of the force network.  The higher moments $\mathbb{S}_{2n}$ give higher order force-orientational information.
Considering $\tilde h(\theta)$, and its Fourier coefficients $\mathbb{S}_{2n}$, thus generalizes the stress tensor to higher order  moments.
 
\subsubsection{Uniaxial Compression vs Pure Shearing} 

We will first apply the above to the two cases of uniaxial compression and pure shearing, since they share the same symmetries.  From Figs.~\ref{strains}(a) and \ref{strains}(c), we see that these both have the maximal stress axis in the $\mathbf{\hat x}$ direction, and the minimal stress axis in the $\mathbf{\hat y}$ direction. The reflection symmetry of the deformations $y \leftrightarrow -y$, implies the symmetry $\theta_{ij}\leftrightarrow -\theta_{ij}$, and thus we have $\mathcal{P}(\theta,h)=\mathcal{P}(-\theta,h)$.  Consequently, $\mathcal{P}(\theta)$,  $\tilde h(\theta)$, and $h(\theta)$ are all symmetric about $\theta=0$, and so in the plots below we show results restricted to the range $\theta\in[0,\pi/2)$.

In Figs.~\ref{P-vs-theta}(a) and \ref{P-vs-theta}(b) we plot $\mathcal{P}(\theta)$ vs $\theta$ for several different packing fractions $\phi$ at the strain rate $\dot\epsilon=10^{-7}$, for uniaxial compression and pure shearing respectively.  In Figs.~\ref{tilde_h-vs-theta}(a) and \ref{tilde_h-vs-theta}(b) we similarly plot the corresponding $\tilde h(\theta)/\langle h\rangle$; we normalize $\tilde h(\theta)$ by $\langle h\rangle$ so that all curves have a common average of $1/\pi$.  We use a common scale for the vertical axes of both the uniaxial and pure shear cases,  so as to allow an easy visual comparison between the two.
From Figs.~\ref{P-vs-theta} and \ref{tilde_h-vs-theta} one see that the anisotropy decreases as one approaches $\phi_J\approx 0.84$.  Pure shearing results in greater anisotropy than uniaxial compression.  The anisotropy of the contact network, given by $\mathcal{P}(\theta)$, involves larger, higher order, Fourier components than does the force network, given  by $\tilde h(\theta)$, particularly for pure shearing.

\begin{figure}
\centering
\includegraphics[width=3.4in]{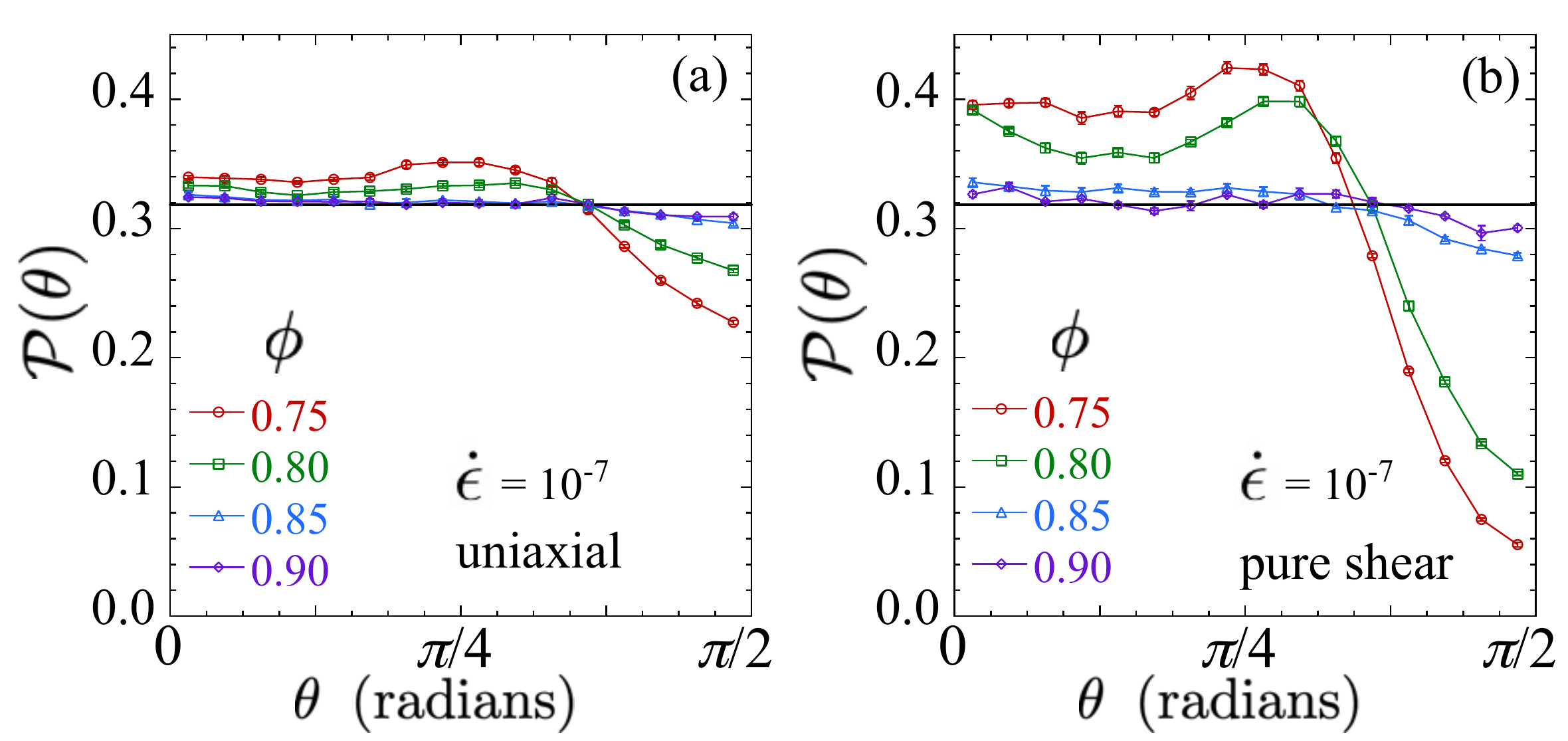}
\caption{
Probability density $\mathcal{P}(\theta)$ for the system contact network to have a bond directed at angle $\theta$ with respect to the compressive direction $\mathbf{\hat x}$.  Results are plotted vs $\theta$ for several different packing fractions $\phi$ at strain rate $\dot\epsilon=10^{-7}$.  (a) is for uniaxial compression, while (b) is for pure shearing.  The solid horizontal black line at $\mathcal{P}=1/\pi$ represents the average value.  The system has $N=32768$ particles.
}
\label{P-vs-theta}
\end{figure}

\begin{figure}
\centering
\includegraphics[width=3.4in]{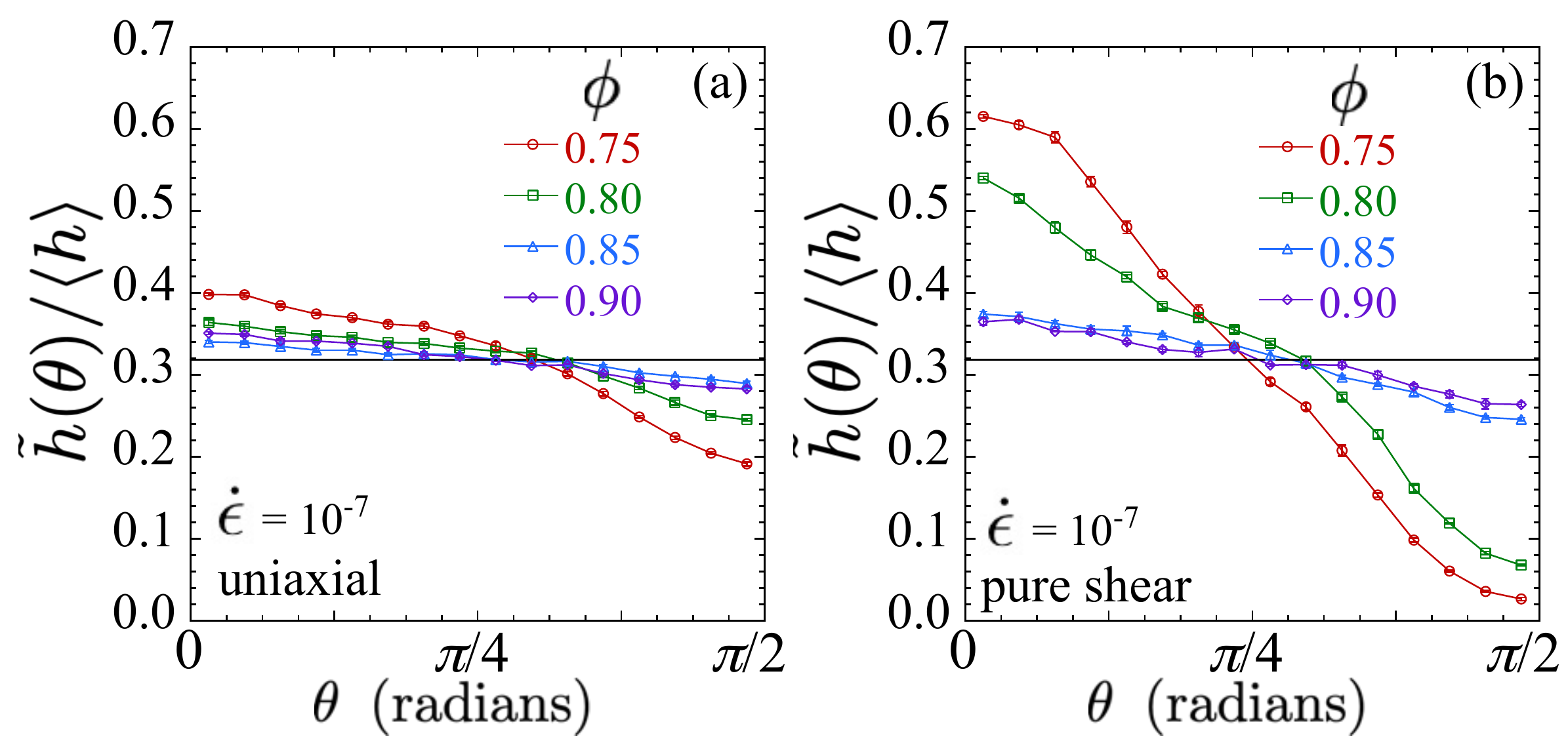}
\caption{Average force-moment $h=f_{ij}^\mathrm{el}\,r_{ij}$ per radian at contact bond angle $\theta$, $\tilde h(\theta)$, normalized by the average force moment $\langle h\rangle$, vs $\theta$ for  several different packing fractions $\phi$ at strain rate $\dot\epsilon=10^{-7}$.  (a) is for uniaxial compression, while (b) is for pure shearing.  The solid horizontal black line at $\tilde h/\langle h\rangle=1/\pi$ represents the average value.  The system has $N=32768$ particles.
}
\label{tilde_h-vs-theta}
\end{figure}

In Figs.~\ref{h-vs-theta}(a) and \ref{h-vs-theta}(b) we plot $h(\theta)/\pi\langle h\rangle$ vs $\theta$ 
for the same parameters as in Figs.~\ref{P-vs-theta} and \ref{tilde_h-vs-theta}.
We normalize $h(\theta)$ by $\pi\langle h\rangle$ so that all curves have the same average $1/\pi$ as the $\tilde h(\theta)$ curves in Fig.~\ref{tilde_h-vs-theta}.  Comparing to $\tilde h(\theta)$ we see that $h(\theta)$ has a somewhat smaller anisotropy, yet the anisotropy in the force-moments remains sizeable.  As $\phi \to \phi_J$, we see, as might be expected, that the forces are greater than average for $0\le \theta\lesssim \pi/4$, and less than average for $\pi/4\lesssim\theta\le\pi/2$.

\begin{figure}
\centering
\includegraphics[width=3.4in]{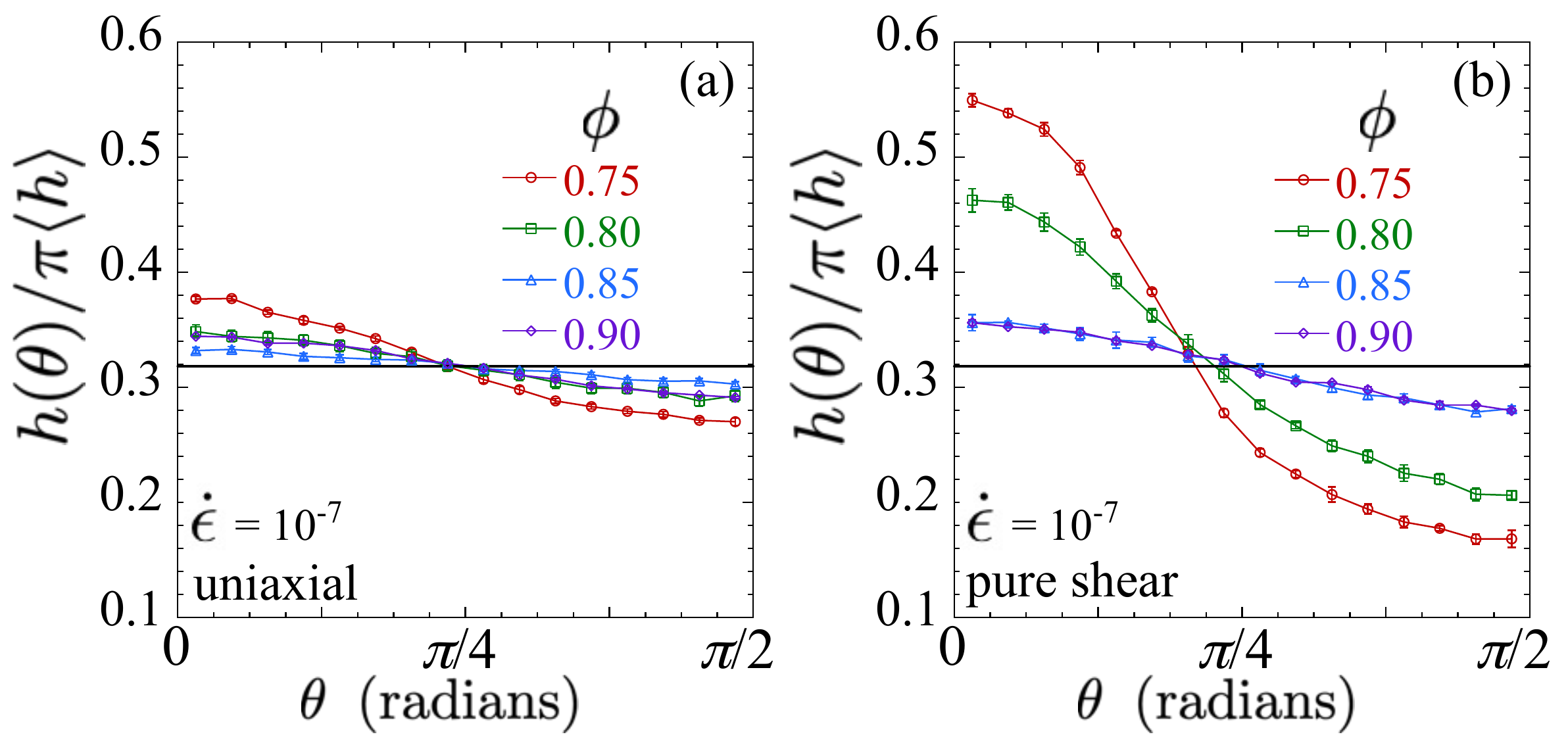}
\caption{Average force-moment on a contact bond at angle $\theta$, $h(\theta)=\tilde h(\theta)/\mathcal{P}(\theta)$, normalized by $\pi \langle h\rangle$, vs $\theta$ for several different packing fractions $\phi$ at strain rate $\dot\epsilon=10^{-7}$.  (a) is for uniaxial compression, while (b) is for pure shearing.  The solid horizontal black line at $\tilde h/\pi \langle h\rangle=1/\pi$ represents the average value.
The system has $N=32768$ particles.
}
\label{h-vs-theta}
\end{figure}

We now quantify the trends seen in Figs.~\ref{P-vs-theta} and \ref{tilde_h-vs-theta} by computing the orientational order parameters $(S_{2n},\theta_{2n})$ and $(\mathbb{S}_{2n},\vartheta_{2n})$.  Because of the symmetry $\theta\leftrightarrow -\theta$, we must have
$B_n=\langle \sin 2n\theta\rangle =0$ and $D_n=\langle h \sin 2n\theta\rangle =0$.  From Eqs.~(\ref{eS2n}) and (\ref{eSp2n}) we therefore have, 
\begin{equation}
S_{2n}=|\langle\cos 2n\theta\rangle|\quad\text{and}\quad\mathbb{S}_{2n}=|\langle h\cos 2n\theta\rangle|/\langle h\rangle,
\end{equation}
and from Eqs.~(\ref{eth2n}) and (\ref{ethp2n}) we have for the orientations,
\begin{equation}
\tan (2n\theta_{2n})=0\quad\text{and}\quad\tan (2n\vartheta_{2n}) =0.
\end{equation}
We therefore have $\theta_{2n}=0$ when $\langle\cos 2n\theta\rangle >0$, and $\theta_{2n}=\pi/2n$ when $\langle \cos 2n\theta\rangle <0$, and similarly for $\vartheta_{2n}$.

Because $\theta_{2n}$ and $\vartheta_{2n}$ are restricted to only these two possible values, we will drop the absolute value sign in the definitions of $S_2$ and $\mathbb{S}_{2n}$ and henceforth, for uniaxial compression and pure shear, adopt the notation,
\begin{equation}
S_{2n}\equiv\langle\cos2n\theta\rangle,
\quad\left\{
\begin{array}{ll}
\theta_{2n}=0&\text{when $S_{2n}>0$}\\
\theta_{2n}=\pi/2n&\text{when $S_{2n}<0$.}
\end{array}
\right.
\label{eS2}
\end{equation}
and
\begin{equation}
\mathbb{S}_{2n}\equiv\dfrac{\langle h\cos2n\theta\rangle}{\langle h\rangle},
\quad\left\{
\begin{array}{ll}
\vartheta_{2n}=0&\text{when $\mathbb{S}_{2n}>0$}\\
\vartheta_{2n}=\pi/2n&\text{when $\mathbb{S}_{2n}<0$.}
\end{array}
\right.
\label{eSp2}
\end{equation}

In Fig.~\ref{S2n-vs-phi-1234}(a) we plot the order parameters for the contact network, $S_{2n}$ vs $\phi$, for $n=1,2,3,$ and 4, at the fixed strain rate $\dot\epsilon =10^{-7}$. 
Closed symbols represent pure shear, while open symbols give uniaxial compression.  The $n=1$ nematic order parameter $S_2$ is the same as the fabric anisotropy $\Delta\lambda$ previously shown in Fig.~\ref{mu-lambda-vs-phi}(b).  We see that the $n=2$ tetratic order parameter $S_4$ is comparable in size to the $n=1$ nematic order, $|S_4|\approx |S_2|$, while the $n=3$ hexatic ordering $S_6$ is noticeable but smaller. $S_8$ and higher order terms are generally quite small. 
That $S_2,S_6>0$ indicates that the nematic and hexatic orderings are oriented at $\theta_2,\theta_6=0$, while $S_4<0$ means that the tetratic ordering is at $\theta_4 = \pi/4$, along the diagonal.  This tetratic ordering is responsible for the shoulder seen in $\mathcal{P}(\theta)$ at $\theta=\pi/4$ in Fig.~\ref{P-vs-theta}.
These results indicate the expected conclusion that bonds prefer to orient along the compressive direction $\mathbf{\hat x}$, and are least likely to orient along the transverse direction $\mathbf{\hat y}$ {\color{black}\cite{Behringer}}.
As noted earlier for the nematic ordering, we see that  for all  moments the orientational ordering of pure shearing is greater   than for uniaxial compression, $|S_{2n}^\mathrm{ps}|>|S_{2n}^\mathrm{uni}|$.

In Fig.~\ref{S2n-vs-phi-1234}(b) we show the corresponding plot of the order parameters for the force network, $\mathbb{S}_{2n}$.
$\mathbb{S}_2$ is the same as $\mu$ previously shown in Fig.~\ref{mu-lambda-vs-phi}(a).  Comparing to the $S_{2n}$ from the contact network, we see that  $\mathbb{S}_{2n}$ is generally smaller than $S_{2n}$ for $n>1$, and thus the dominant mode of anisotropy in the force network is from the $\mathbb{S}_2=\mu$ nematic term.  This indicates that weighting the contact bonds by their force moment $h_{ij}$ serves to reduce the non-nematic components of the anisotropy present in the contact network geometry.  As with $S_{2n}$, we see that $\mathbb{S}_4$ is generally negative while $\mathbb{S}_6$ is positive.

\begin{figure}
\centering
\includegraphics[width=3.4in]{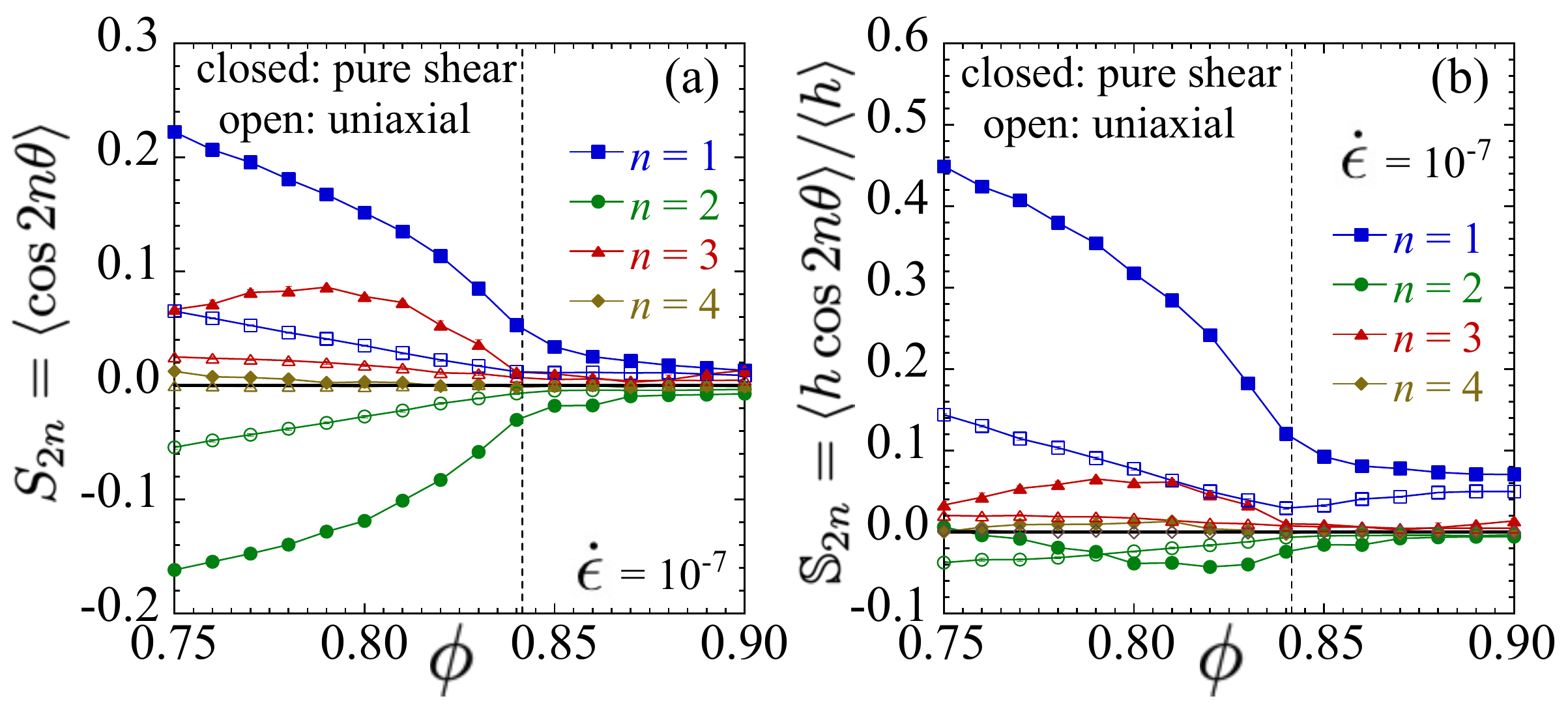}
\caption{(a) Orientational order parameter $S_{2n}$ for the contact network, and (b)  orientational order parameter $\mathbb{S}_{2n}$ for  the force network, vs $\phi$ for strain rate $\dot\epsilon=10^{-7}$.  Results are shown for nematic ($n=1$),  tetratic ($n=2$),  hexatic ($n=3$), and 8-fold ($n=4$) orientational order.  $S_2=\Delta\lambda$ is the same as the fabric tensor anisotropy, while $\mathbb{S}_2=\mu$ is the same as the stress tensor anisotropy.  Closed symbols denote pure shearing while open symbols denote uniaxial compression.  
The vertical dashed line indicates the isotropic compression-driven jamming $\phi_J = 0.8415$. The system has $N=32768$ particles.
}
\label{S2n-vs-phi-1234}
\end{figure}

From our above results, shown in Figs.~\ref{mu-lambda-vs-phi} and \ref{S2n-vs-phi-1234}, it is clear that the difference in anisotropy, comparing the two cases of uniaxial compression and pure shearing, is to a great extent due to the difference between the relative magnitudes of the isotropic part to the anisotropic part of the stress and fabric tensors.  It is therefore interesting to subtract off the isotropic part, and to see how only the anisotropic parts compare with each other.  Subtracting off the leading isotropic term from Eqs.~(\ref{eFT}) and (\ref{ehFT}), and normalizing by the magnitude of the nematic term, we consider,
\begin{align}
\dfrac{\Delta\mathcal{P}(\theta)}{S_2} &= \frac{2}{\pi}\sum_{n=1}\dfrac{S_{2n}}{S_2}\cos(2n[\theta-\theta_{2n}])
\label{eDP}\\[10pt]
\dfrac{\Delta\tilde h(\theta)}{\langle h\rangle\mathbb{S}_2} &=  \frac{2}{\pi}\sum_{n=1}\dfrac{\mathbb{S}_{2n}}{\mathbb{S}_2}\cos(2n[\theta-\vartheta_{2n}]).
\label{eDtildeh}
\end{align}

In Figs.~\ref{DP-vs-theta}(a) and \ref{DP-vs-theta}(b) we plot $\Delta\mathcal{P}(\theta)/S_2$ vs $\theta$ for different $\phi$ at strain rate $\dot\epsilon=10^{-7}$, for uniaxial compression and pure shearing, respectively.  In Figs.~\ref{D_tilde_h-vs-theta}(a) and \ref{D_tilde_h-vs-theta}(b) we make similar plots of $\Delta\tilde h(\theta)/\langle h\rangle\mathbb{S}_2$.  In both Figs.~\ref{DP-vs-theta} and \ref{D_tilde_h-vs-theta} the solid black line is the functional form, $(2/\pi)\cos 2\theta$, that one would have if only the nematic $(n=1)$ term was present (since $\theta_2=\vartheta_2=0$).  In Fig.~\ref{DP-vs-theta} we see that $\Delta\mathcal{P}(\theta)$ involves significant higher order terms beyond the nematic, however, qualitatively, there does not appear to be much difference between the two cases of uniaxial compression and pure shearing.  In contrast, Fig.~\ref{D_tilde_h-vs-theta} shows that the nematic term does give a reasonable approximation, and so higher order terms are relatively small.  Again, there is little qualitative differences between uniaxial compression and pure shear.  We thus conclude that there is little difference in the anisotropic parts of either the contact network or the force network, when comparing the two cases of uniaxial compression and pure shearing.  The main difference between these two cases lies in the relative magnitude of the anisotropic term to the isotropic term, i.e., $\Delta\lambda = S_2$ and $\mu=\sigma/p=\mathbb{S}_2$.

\begin{figure}
\centering
\includegraphics[width=3.4in]{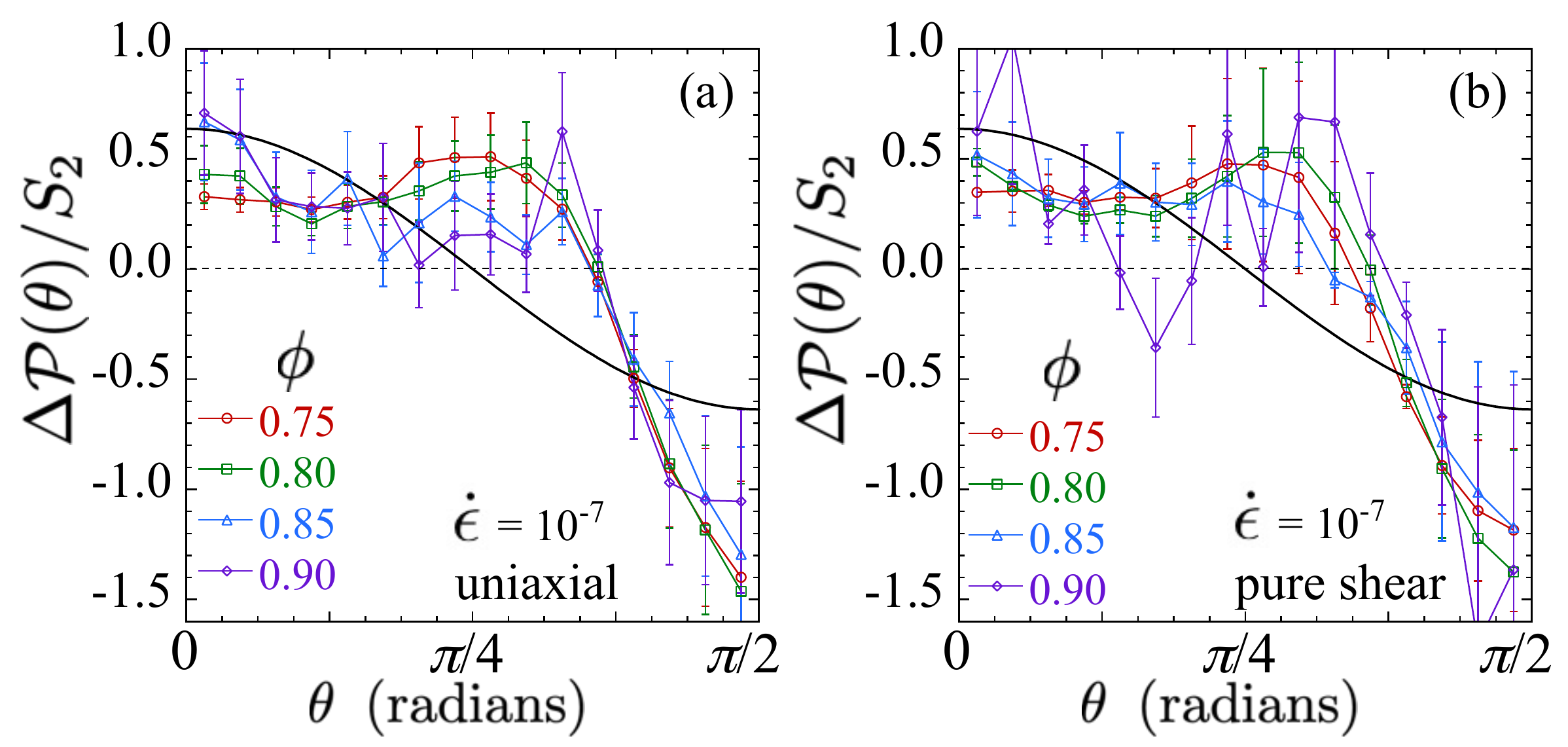}
\caption{Anisotropic part of the contact network bond orientation probability $\Delta\mathcal{P}(\theta)=\mathcal{P}(\theta)-1/\pi$, normalized by the magnitude of the nematic order Fourier coefficient $S_2$, vs $\theta$ for several different packing fractions $\phi$ at strain rate $\dot\epsilon=10^{-7}$.  (a) is for uniaxial compression, while (b) is for pure shearing.  The solid black line gives the functional form for the purely nematic term, $(2/\pi)\cos 2\theta$. 
The system has $N=32768$ particles.
}
\label{DP-vs-theta}
\end{figure}

\begin{figure}
\centering
\includegraphics[width=3.4in]{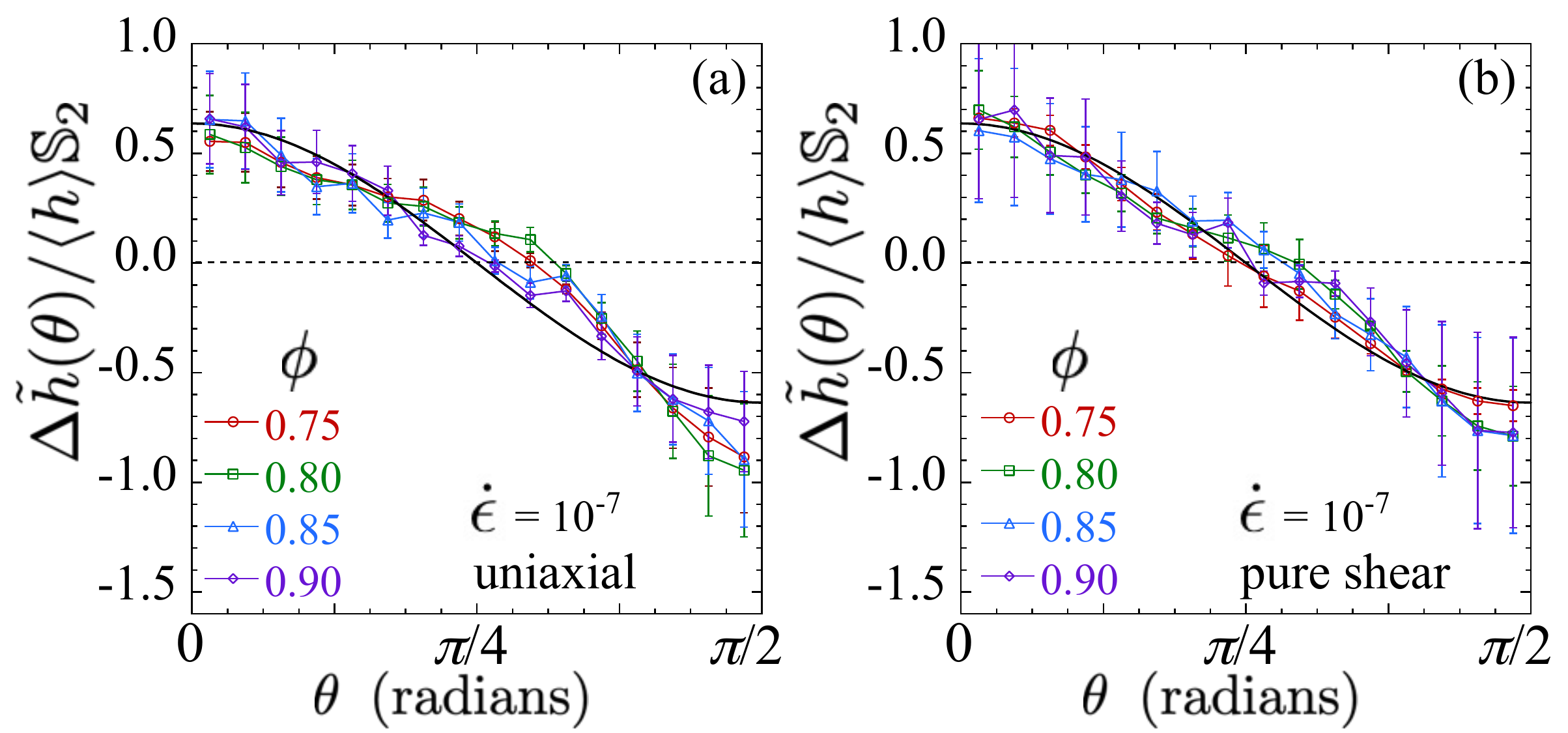}
\caption{Anisotropic part of the average force-moment orientation $\Delta\tilde h(\theta)=\tilde h(\theta)-\langle h\rangle/\pi$, normalized by the magnitude of the nematic order Fourier coefficient $\langle h\rangle\mathbb{S}_2$, vs $\theta$ for several different packing fractions $\phi$ at strain rate $\dot\epsilon=10^{-7}$.  (a) is for uniaxial compression, while (b) is for pure shearing.  The solid black line gives the functional form for the purely nematic term, $(2/\pi)\cos 2\theta$. 
The system has $N=32768$ particles.
}
\label{D_tilde_h-vs-theta}
\end{figure}

To quantify these observations, in Fig.~\ref{An-A1-vs-phi}(a) we plot the ratios $S_{2n}/S_2$ vs $\phi$, at the fixed strain rate $\dot\epsilon=10^{-7}$,  for  tetratic ($n=2$), hexatic ($n=3$), and 8-fold ($n=4$) order.  In in Fig.~\ref{An-A1-vs-phi}(b) we show the corresponding plot for $\mathbb{S}_{2n}/\mathbb{S}_2$.  We see from these plots that there is now relatively little difference between uniaxial compression and pure shearing, and that the magnitudes of these higher order orientational terms are relatively small for the force network at all $\phi$, though not for the contact network.  We thus conclude that the main difference in anisotropy, comparing uniaxial compression with pure shearing, is due to differences in the magnitude of the  nematic ordering.

\begin{figure}
\centering
\includegraphics[width=3.4in]{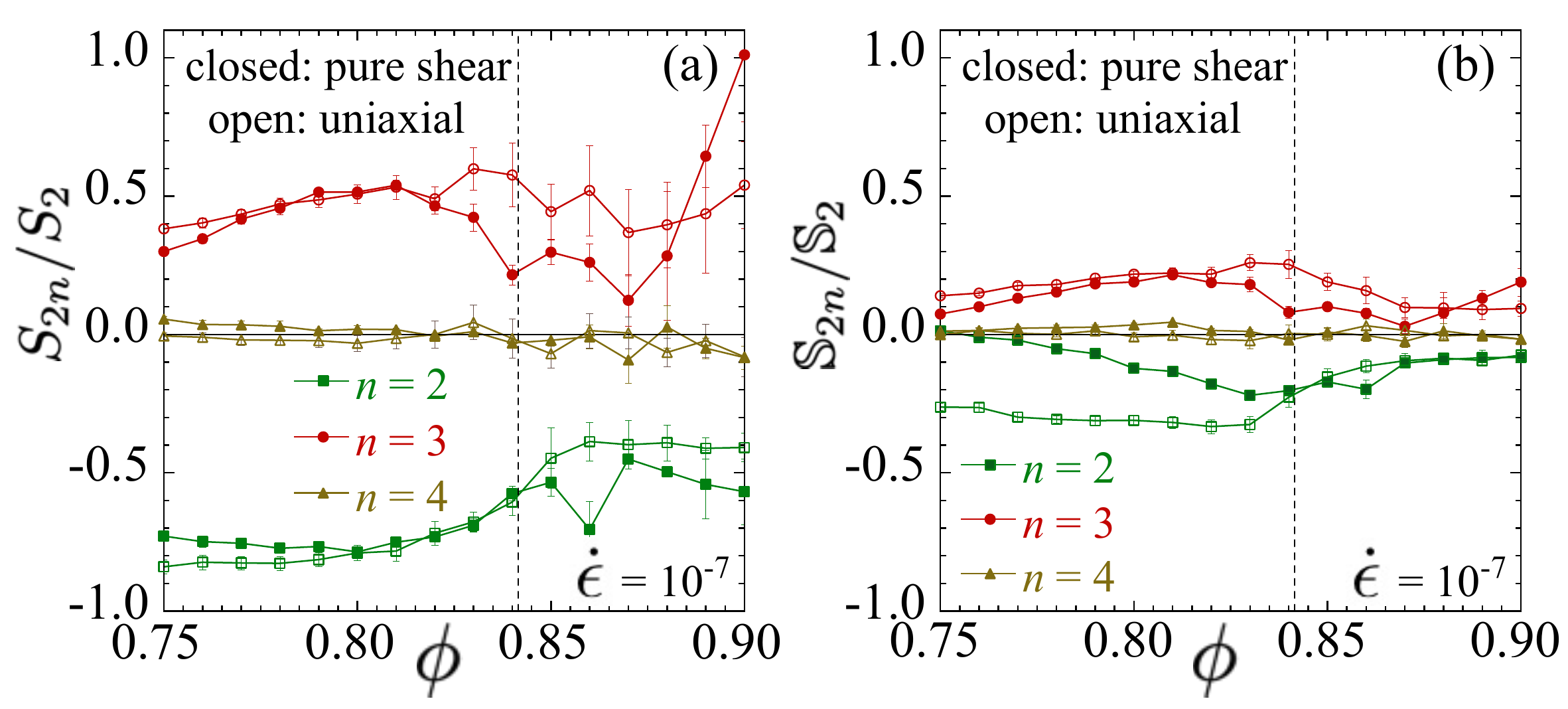}
\caption{(a) Ratio of orientational order parameters $S_{2n}/S_2$ for the bond directions in the geometrical contact  network, and (b) ratio of  orientational order parameters $\mathbb{S}_{2n}/\mathbb{S}_2$ for the  force network, vs $\phi$ for strain rate $\dot\epsilon=10^{-7}$.  Results are shown for  tetratic ($n=2$),  hexatic ($n=3$), and  8-fold ($n=4$) orientational order.  Closed symbols denote pure shearing while open symbols denote uniaxial compression.  
The vertical dashed line indicates the isotropic compression-driven jamming $\phi_J = 0.8415$. The system has $N=32768$ particles.
}
\label{An-A1-vs-phi}
\end{figure}

\subsubsection{Simple Shearing}

We now consider the case of simple shearing.  Unlike uniaxial compression and pure shearing, now there is no symmetry in $\theta\leftrightarrow -\theta$, and so we will show our results for the full range of $\theta\in[-\pi/2,\pi/2)$.  A simple sheared elastic continuum would have its maximal stress axis oriented in the $(1,-1)$ direction.  As we noted in connection with Fig.~ \ref{mu-lambda-vs-phi}, and as we will show explicitly below, for our granular system we do find $\theta_2, \vartheta_2\approx -\pi/4$.

One might therefore think that distributions might be symmetric about $\theta=-\pi/4$.  But we find this is not in general the case.  Unlike uniaxial compression and pure shear, where the orthogonal principle axes of the stress tensor are  the only unique directions in the problem, for simple shear we have as well the average flow direction (in our case $\mathbf{\hat x}$), which may contribute to the orientation of the order parameters $S_{2n}$ and $\mathbb{S}_{2n}$.

In Fig.~\ref{P-vs-theta-ss} we plot $\mathcal{P}(\theta)$ vs $\theta$ for simple shearing, for several different packings $\phi$ at strain rate $\dot\epsilon=10^{-7}$.  As was seen for both uniaxial compression and pure shearing, the degree of anisotropy decreases as the packing $\phi$ increases.
We see a minimum near $\theta=\pi/4$, which might be expected as this is close to the direction of the minimal stress axis.  However, for $\phi<\phi_J^*=0.8435$, we see no maximum at $\phi=-\pi/4$, close to the direction of the maximal stress. There is no symmetry about $\theta_2\approx -\pi/4$.  Instead we see a relatively sharp maximum at $\theta=0$, along the flow direction $\mathbf{\hat x}$, which may be viewed as an analog of the peak at $\theta=\pi/4$ seen in Fig~\ref{P-vs-theta}(b) for pure shearing.
Looking above jamming at $\phi>\phi_J^*=0.8435$, this peak at $\theta=0$ goes away.  Although it is difficult to see in the plot due to the compressed range of $\mathcal{P}(\theta)$ at the larger $\phi$, for $\phi>\phi_J^*$, $\mathcal{P}(\theta)$ does become approximately symmetric about $\theta=-\pi/4$, with a broad maximum at $\theta\approx -\pi/4$, and a sharper minimum at $\theta\approx \pi/4$; the shape of $\mathcal{P}(\theta)$ now looks quite similar to that found for pure shearing in Fig.~\ref{P-vs-theta}(b), only shifted by $-\pi/4$.

\begin{figure}
\centering
\includegraphics[width=3.0in]{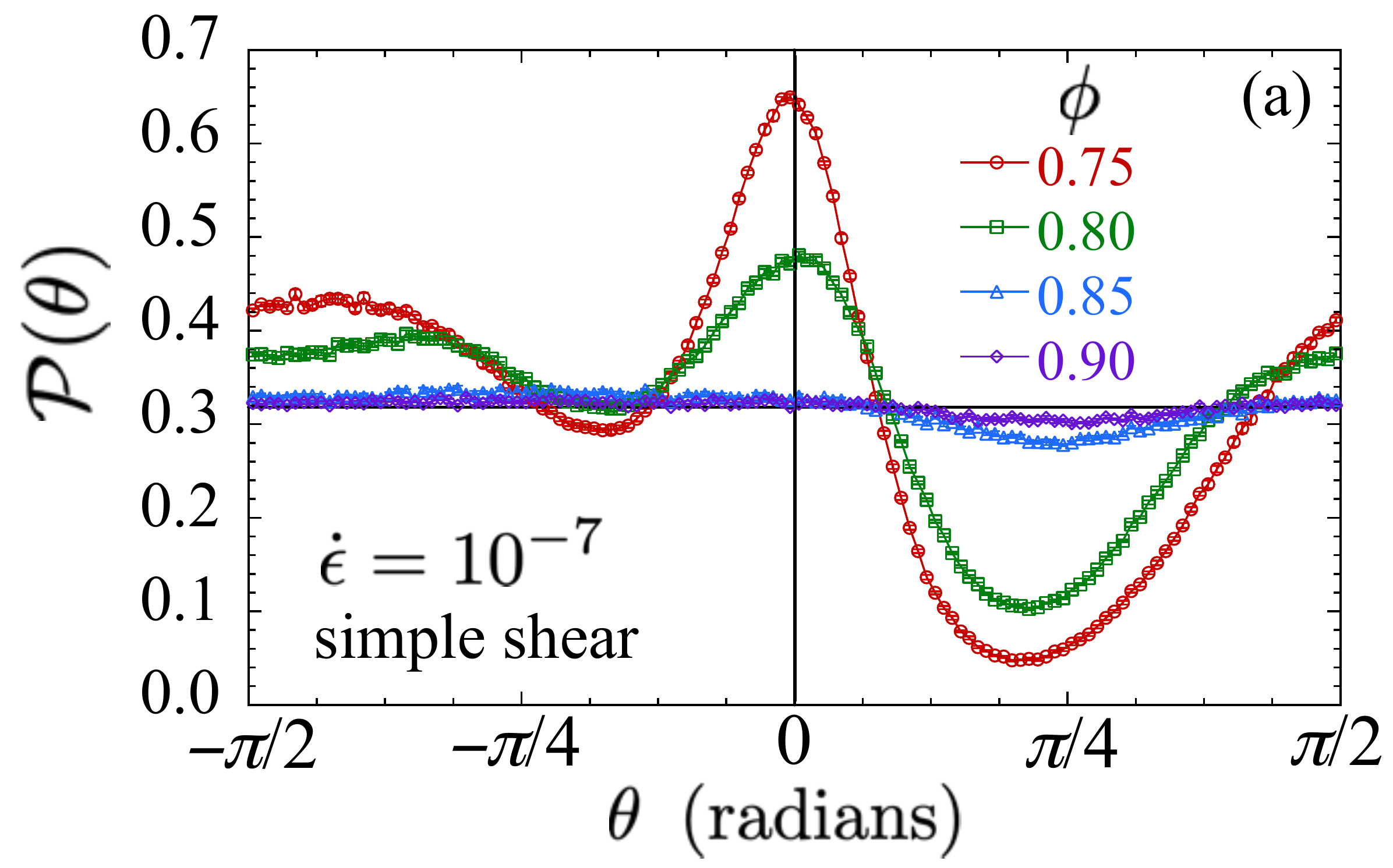}
\caption{Probability density $\mathcal{P}(\theta)$ for the  contact network to have a bond directed at angle $\theta$ with respect to the flow direction $\mathbf{\hat x}$, in simple shearing.  Results are plotted vs $\theta$ for several different packing fractions $\phi$ at strain rate $\dot\epsilon=10^{-7}$.  The solid horizontal black line at $\mathcal{P}=1/\pi$ represents the average value.  The system has $N=32768$ particles.
}
\label{P-vs-theta-ss}
\end{figure}

In Fig.~\ref{tilde_h-vs-theta-ss} we  show the corresponding plot of the force-moment per radian, $\tilde h(\theta)$.  The shape of $\tilde h(\theta)$ is similar to that of $\mathcal{P}(\theta)$, except  there is now a peak just below $\theta=-\pi/4$ from the large forces at the contacts along the maximal stress direction.  In Fig.~\ref{h-vs-theta-ss} we show $h(\theta)=\tilde h(\theta)/\mathcal{P}(\theta)$, which measures the average value of the force-moment on bonds at angle $\theta$, independent of the probability for there to be a bond at $\theta$.  With $\mathcal{P}(\theta)$ factored out, the behavior of $h(\theta)$ is more easily understood: force-moments are largest along the maximal stress direction at $\theta=-\pi/4$, and smallest along the minimal stress direction at $\theta=\pi/4$.  The distribution $h(\theta)$ is symmetric about its maximum for all packings $\phi$. For $\phi=0.85$ and 0.90, above jamming, $h(\theta)/\langle h\rangle$ are essentially equal.

\begin{figure}
\centering
\includegraphics[width=3.0in]{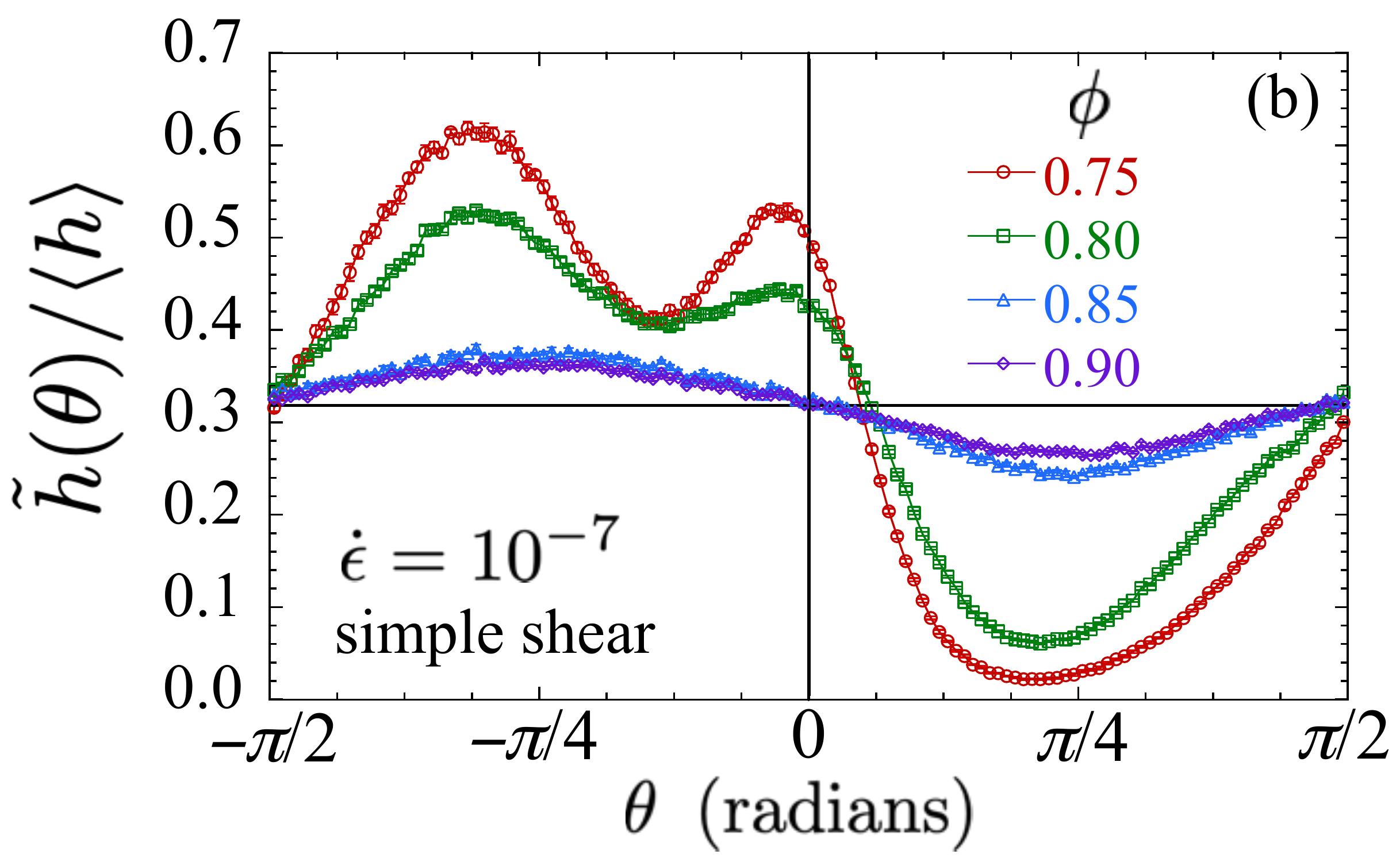}
\caption{Average force-moment per radian $\tilde h(\theta)$, normalized by the average force moment $\langle h\rangle$, in simple shearing.  Results are plotted  vs bond angle $\theta$ for  several different packing fractions $\phi$ at strain rate $\dot\epsilon=10^{-7}$.  The solid horizontal black line at $\tilde h/\langle h\rangle=1/\pi$ represents the average value.  The system has $N=32768$ particles.}
\label{tilde_h-vs-theta-ss}
\end{figure}

\begin{figure}
\centering
\includegraphics[width=3.0in]{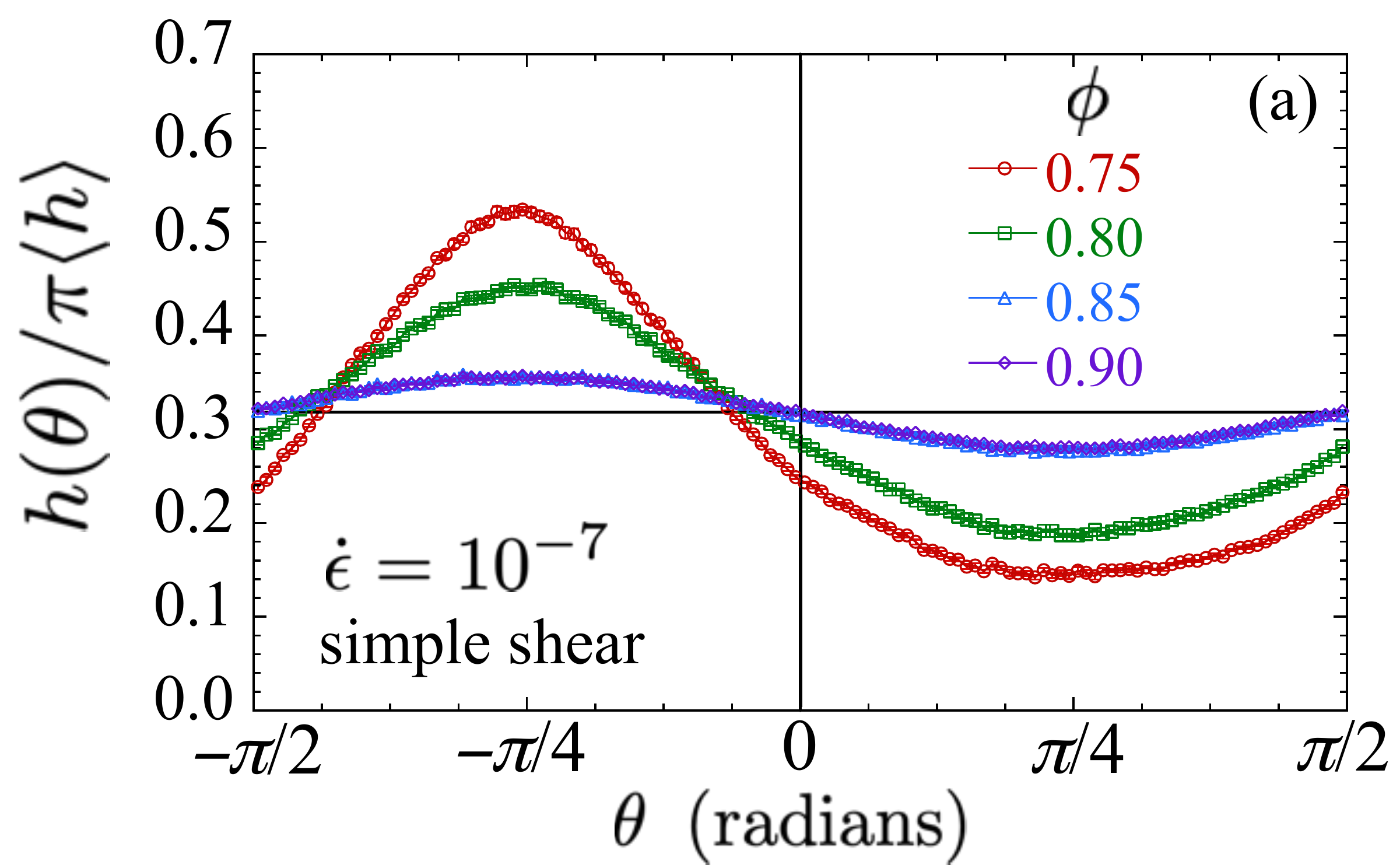}
\caption{Average force-moment $h(\theta)=\tilde h(\theta)/\mathcal{P}(\theta)$, normalized by $\pi \langle h\rangle$, in simple shearing.  Results are plotted vs bond angle $\theta$ for several different packing fractions $\phi$ at strain rate $\dot\epsilon=10^{-7}$.  The solid horizontal black line at $\tilde h/\pi \langle h\rangle=1/\pi$ represents the average value.
The system has $N=32768$ particles.
}
\label{h-vs-theta-ss}
\end{figure}

To make the above observations more quantitative, in Figs.~\ref{S2n-vs-phi-1234-ss}(a) and \ref{S2n-vs-phi-1234-ss}(b) we plot the magnitude of the contact network and force network orientational order parameters $S_{2n}$ and $\mathbb{S}_{2n}$ vs $\phi$ for $n=1$, 2, 3, and 4, at the fixed strain rate $\dot\epsilon=10^{-7}$, and compare those against the corresponding values for pure shearing.  
Note, here we take $S_{2n}$ and $\mathbb{S}_{2n}$ to be strictly positive, rather than use the sign convention of Eqs.~(\ref{eS2}) and (\ref{eSp2}), since for simple shear there is no a priori restriction on the values of $2n\theta_{2n}$ and $2n\vartheta_{2n}$ to just two values as in the case of pure shear and uniaxial compression.

From Fig.~\ref{S2n-vs-phi-1234-ss}(a) we see, as found previously in Fig.~\ref{mu-lambda-vs-phi}(b), that when comparing simple and pure shearing, the $n=1$ nematic terms $S_2$ are roughly equal for the whole range of $\phi$ shown.  However, looking at the higher moments, we see that $S_{2n}$  for simple and pure shearing are essentially equal only once $\phi\gtrsim 0.82$.  For $\phi\lesssim 0.82$ there is a pronounced difference.  As $\phi$ decreases, the difference in orientational ordering between simple and pure shearing, increases. From Fig.~\ref{S2n-vs-phi-1234-ss}(b) we see that the same is true for the  $\mathbb{S}_{2n}$ of the force network.  It is interesting to note that, for the contact network, $S_4$ for the $n=2$ tetratic order is greater than $S_2$ for the $n=1$ nematic order, at the smaller packings $\phi$.  This is related to the strong peak in $\mathcal{P}(\theta)$ at $\theta=0$, seen in Fig.~\ref{P-vs-theta-ss}.  As was true for both pure shearing and uniaxial compression, we find that the higher order moments ($n>1$)  for the force network, $\mathbb{S}_{2n}$, are smaller than the corresponding contact network moments, $S_{2n}$, when measured relative to the  $n=1$ moment.  The variation of the force-moments with bond direction in the force network tends to suppress the higher order moments of anisotropy as compared to the purely geometric contact network.

In Fig.~\ref{theta-thetah-vs-phi-1234} we show the  angles $\theta_{2n}$ and $\vartheta_{2n}$ of the orientation order parameters for the contact and force networks.  Results are plotted vs $\phi$ for simple shearing at strain rate $\dot\epsilon=10^{-7}$.  We see, as mentioned before, that the nematic order is oriented at $\theta_2,\vartheta_2\approx -\pi/4$.  The tetratic order is oriented at $\theta_4,\vartheta_4\approx 0$, though for the force network $\vartheta_4$ increases as $\phi$ decreases.  Comparing to pure shear, where $\theta_2^\mathrm{ps},\vartheta_2^\mathrm{ps}=0$ and $\theta_4^\mathrm{ps},\vartheta_4^\mathrm{ps}=\pi/4$, we see that our results for simple shear, $\theta_2^\mathrm{ss},\vartheta_2^\mathrm{ss}\approx -\pi/4$ and $\theta_4^\mathrm{ss},\vartheta_4^\mathrm{ss}\approx 0$, represent a simple clockwise  rotation of $S_2, \mathbb{S}_2$ and $S_4,\mathbb{S}_4$  by $\pi/4$ when going from pure shear to simple shear, the same rotation as for the principle stress axes.  However, no such simple explanation applies to the higher moments, where the orientations of $S_6,\mathbb{S}_6$ and $S_8,\mathbb{S}_8$ for simple shearing seem to have no clear relation to those for pure  shearing.

\begin{figure}
\centering
\includegraphics[width=3.4in]{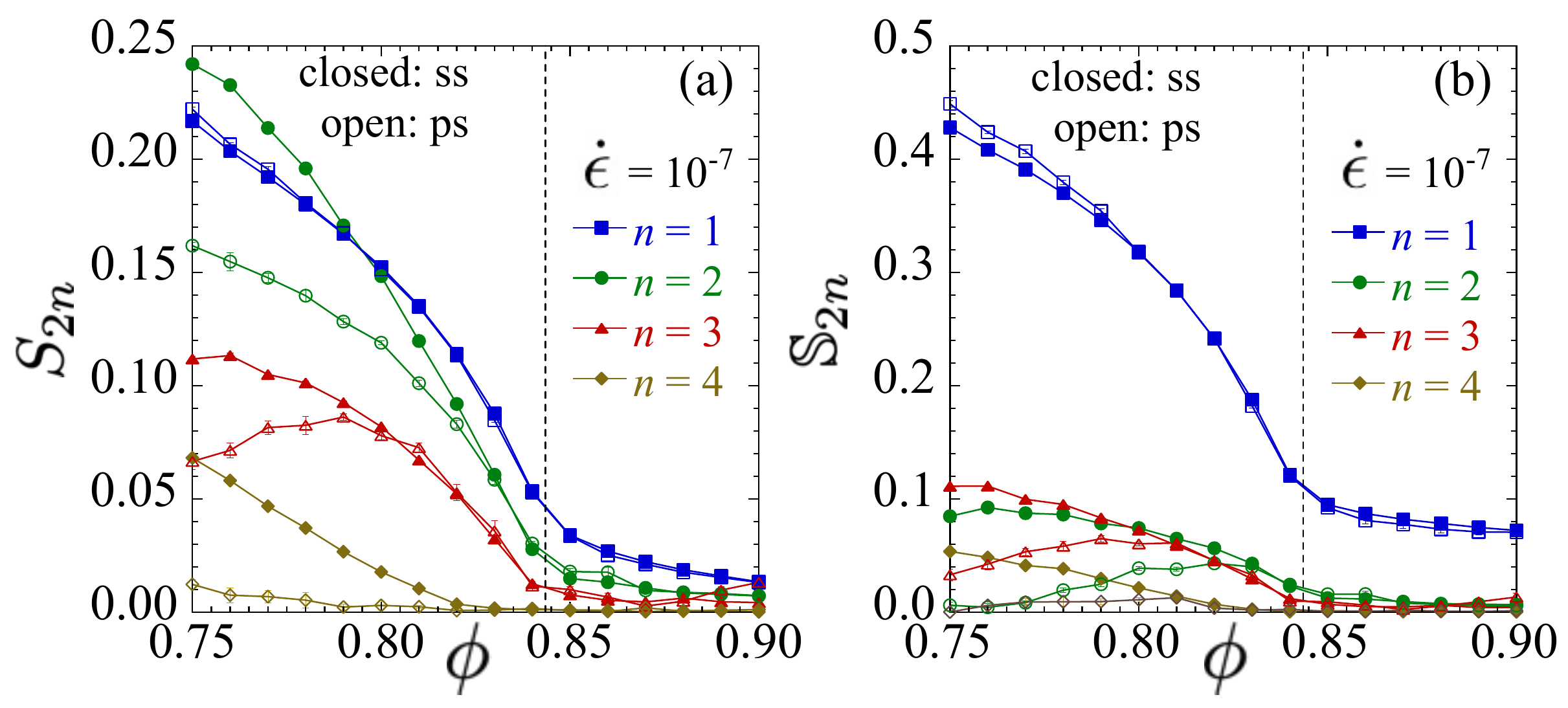}
\caption{
(a) Orientational order parameter $S_{2n}$ for the contact network, and (b)  orientational order parameter $\mathbb{S}_{2n}$ for  the force network, for simple shearing (closed symbols) compared to pure shearing (open symbols).  Results are plotted vs the packing $\phi$ for strain rate $\dot\epsilon=10^{-7}$, showing nematic ($n=1$),  tetratic ($n=2$),  hexatic ($n=3$), and 8-fold ($n=4$) orientational order.  The vertical dashed line indicates the shear-driven jamming $\phi_J = 0.8435$. The system has $N=32768$ particles.
}
\label{S2n-vs-phi-1234-ss}
\end{figure}

\begin{figure}
\centering
\includegraphics[width=3.in]{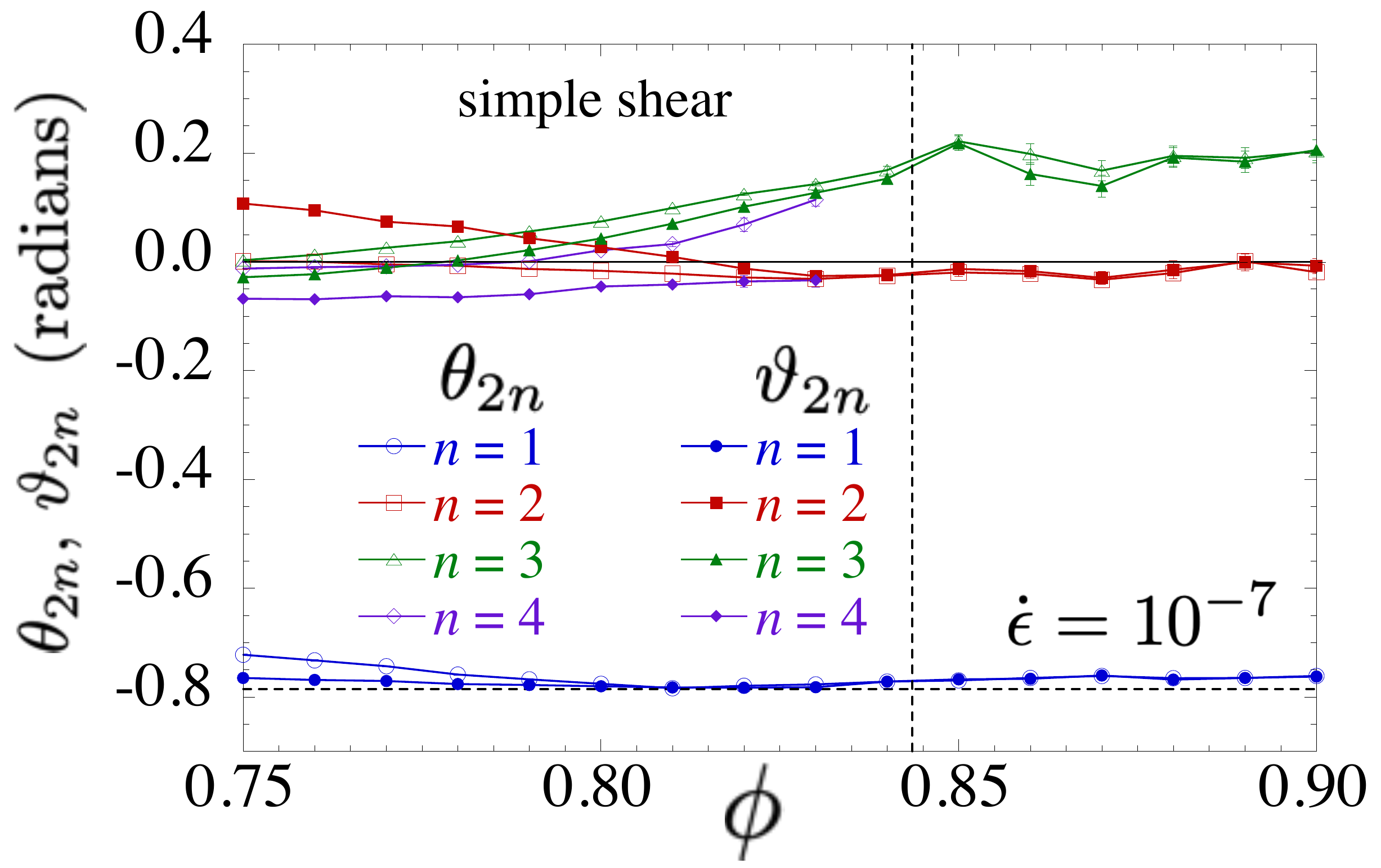}
\caption{Orientation angles $\theta_{2n}$ and $\vartheta_{2n}$ of the order parameters $S_{2n}$ and $\mathbb{S}_{2n}$ for simple shearing.  Results are plotted vs the packing $\phi$ for strain rate $\dot\epsilon=10^{-7}$.  The vertical dashed line represents the shear-driven jamming $\phi_J^*=0.8435$.  The horizontal dashed line is at $\theta=-\pi/4$.  For the 8-fold ordering ($n=4$) we do not show results for $\phi>0.83$ since then $S_8$ and $\mathbb{S}_8$ are too small to determine $\theta_8$ and $\vartheta_8$ reliably.
The system has $N=32768$ particles.
}
\label{theta-thetah-vs-phi-1234}
\end{figure}

Finally, as we did in Figs.~\ref{DP-vs-theta} and \ref{D_tilde_h-vs-theta} for pure shear and uniaxial compression, we can look at just the anisotropic parts of $\mathcal{P}(\theta)$ and $\tilde h(\theta)$ for simple shear.  In Fig.~\ref{DP-vs-theta-ss} we plot $\Delta\mathcal{P}(\theta)/S_2$, defined in Eq.~(\ref{eDP}).  In Fig.~\ref{D_tilde_h-vs-theta-ss} we plot $\Delta\tilde h(\theta)/\langle h\rangle \mathbb{S}_2$, defined in Eq.~(\ref{eDtildeh}). Results are plotted vs $\theta$ for several different packings $\phi$ at the strain rate $\dot\epsilon=10^{-7}$.  In both figures, the solid black curve represents $(2/\pi)\cos 2(\theta+\pi/4)$, which is what we would have if only the nematic term were present (here we take $\theta_2,\vartheta_2 = -\pi/4$).  As was seen for pure shear and uniaxial compression, we find also for simple shear that the contact network $\Delta\mathcal{P}(\theta)/S_2$ retains significant higher order moments even as one goes above $\phi_J^*$, whereas the force network $\Delta\tilde h/\langle h\rangle \mathbb{S}_2$ becomes well described by just the nematic term.

\begin{figure}
\centering
\includegraphics[width=3.in]{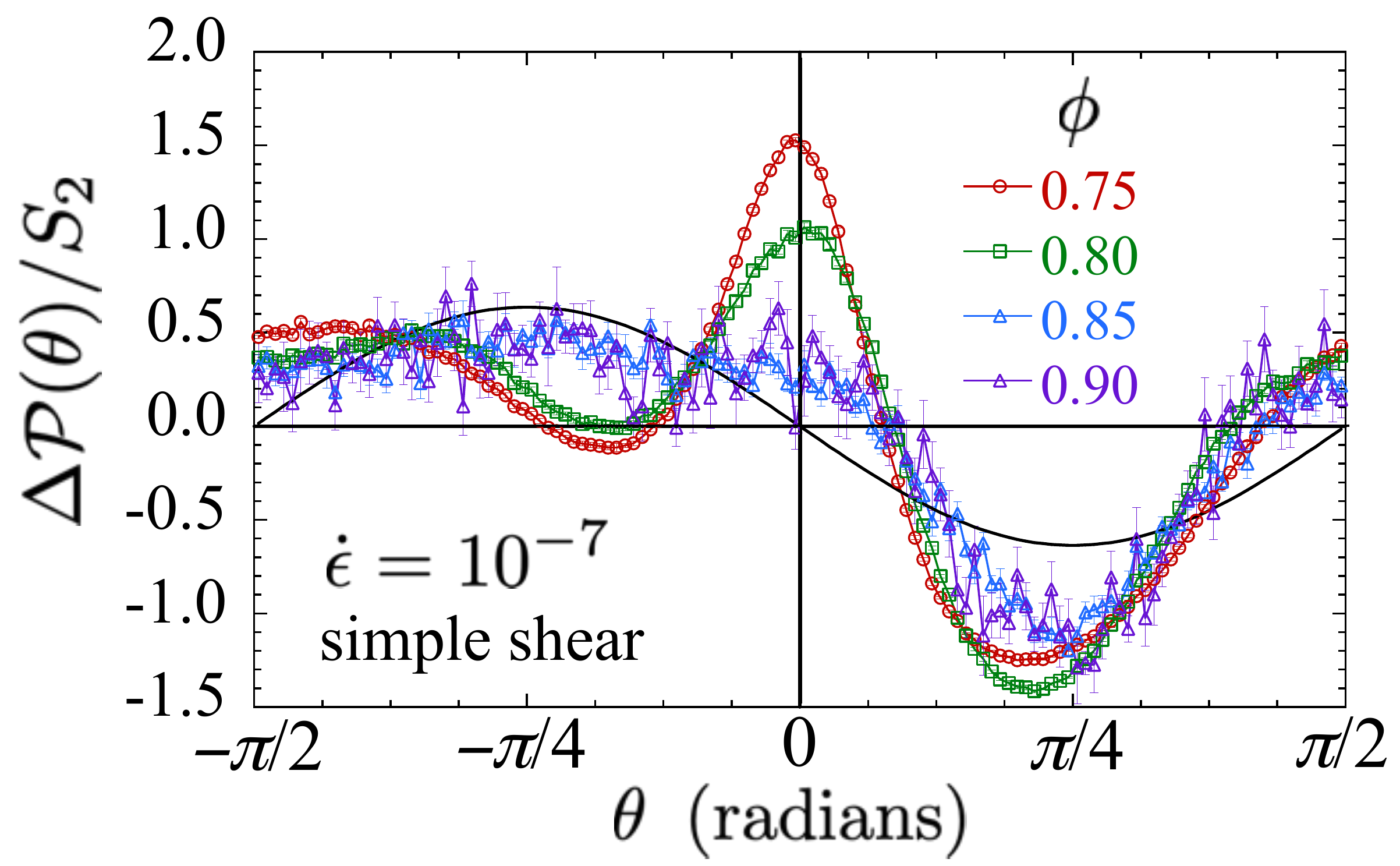}
\caption{For simple shearing:  Anisotropic part of the contact network bond orientation probability $\Delta\mathcal{P}(\theta)=\mathcal{P}(\theta)-1/\pi$, normalized by the magnitude of the nematic order Fourier coefficient $S_2$, vs $\theta$ for several different packing fractions $\phi$ at strain rate $\dot\epsilon=10^{-7}$.  The solid black line gives the functional form for the purely nematic term, $(2/\pi)\cos 2(\theta-\theta_2)$, where we take $\theta_2=-\pi/4$.
The system has $N=32768$ particles.
}
\label{DP-vs-theta-ss}
\end{figure}

\begin{figure}
\centering
\includegraphics[width=3.in]{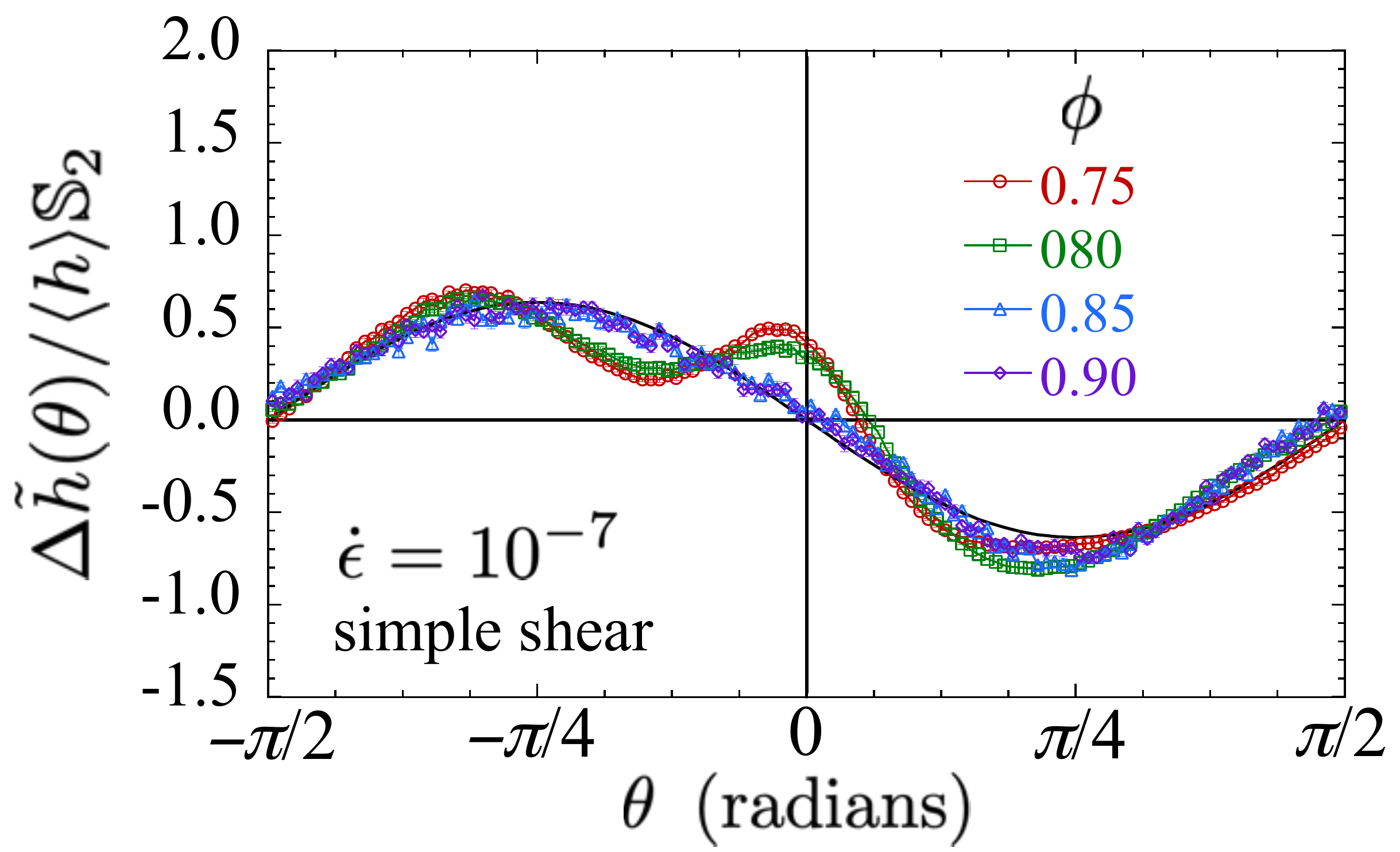}
\caption{For simple shearing:  Anisotropic part of the average force-moment orientation $\Delta\tilde h(\theta)=\tilde h(\theta)-\langle h\rangle/\pi$, normalized by the magnitude of the nematic order Fourier coefficient $\langle h\rangle\mathbb{S}_2$, vs $\theta$ for several different packing fractions $\phi$ at strain rate $\dot\epsilon=10^{-7}$.   The solid black line gives the functional form for the purely nematic term, $(2/\pi)\cos 2(\theta-\vartheta_2)$, where we take $\vartheta_2=-\pi/4$.
The system has $N=32768$ particles.
}
\label{D_tilde_h-vs-theta-ss}
\end{figure}

}

\section{Summary}
\label{sum}

We have carried out numerical simulations of athermal, frictionless, overdamped,  bidisperse circular disks in two dimensions, within a simple model for a non-Brownian suspension, as the packing fraction is increased though the jamming transition.  We compare the stresses that result when the system is deformed by isotropic compression, uniaxial compression,  pure shearing, and {\color{black}simple shearing}, all applied at a fixed strain rate $\dot\epsilon$.

Below jamming, the pressure $p$ arising from uniaxial compression is found to be roughly equal  that of isotropic compression, while the pressure from pure shearing {\color{black} is roughly equal to that of simple shearing.  However the shearing pressure is about an order of magnitude smaller than that from compression.}  Above jamming, all {\color{black}four} cases approach roughly the same pressure {\color{black}as the strain rate $\dot\epsilon$ decreases}.  The shear stress $\sigma$ for isotropic compression is, by symmetry, equal to zero.  {\color{black}The shear stress of pure shearing is roughly equal to that of simple shearing, while the}  shear stress for uniaxial compression is greater than that of  pure/simple shearing below jamming, {\color{black}but smaller than pure/simple shearing above jamming. However the shear stress from uniaxial compression is of the same order of magnitude as the other cases.}  

By comparing the stress ratios of the {\color{black}four} types of deformation we have argued in Fig.~\ref{Rp-Rs-vs-phi}(a)  that isotropic compression and uniaxial compression have the same jamming packing $\phi_J$, with bulk viscosities $\zeta=p/\dot\gamma$ that diverge with the same critical exponent $\beta$.  {\color{black}Similarly, in Fig.~\ref{Rp-Rs-vs-phi}(b) we argued that pure and simple shearing have the same jamming packing and critical exponent.  
However, in Figs.~\ref{Rp-Rs-vs-phi}(c) and (d) we argued that} the jamming packing for  shearing $\phi_J^*$  is slightly larger than the $\phi_J$ for compression.  In Appendix B we provide a detailed critical scaling analysis of our pure shearing results that finds that the pure shearing $\phi_J^*$ is indeed  greater than $\phi_J$, and that this $\phi_J^*$ is equal to the jamming packing previously found for simple shearing \cite{OT2}.  We further find that pure shearing has the same critical exponents, for example $\beta$, as previously  found for compression \cite{PeshkovTeitel} and for simple shearing \cite{OT2}.  Thus stress-isotropic jamming is in the same critical universality class as stress-anisotropic jamming in two dimensions.

The strain rate tensor for uniaxial compression can be viewed as a superposition of the strain rate tensors for isotropic compression plus pure shearing, $\dot{\boldsymbol{\epsilon}}_\mathrm{uni}=\dot{\boldsymbol{\epsilon}}_\mathrm{iso}+\dot{\boldsymbol{\epsilon}}_\mathrm{ps}$.  We have therefore asked if there is any similar superposition for the resulting stresses in the linear rheology region below jamming.   Our conclusion is that there is no such superposition for stresses.

Finally, {\color{black}we have considered the three deformations that result in an anisotropic stress tensor, uniaxial compression, pure shearing, and simple shearing, and}
compared the anisotropy of {\color{black} the corresponding} configurations, {\color{black} considering both the contact network of  bonds and the force network of bonds weighted by their force-moment}.  We have considered both the stress tensor anisotropy $\mu=\sigma/p$ and the anisotropy of the fabric tensor $\Delta\lambda$ of the contact network.  Both parameters approach a finite limiting curve as $\dot\epsilon\to 0$, demonstrating that the systems remain anisotropic both at jamming and above.
We find that $\mu$ behaves qualitatively the same, as a function of packing $\phi$ and strain rate $\dot\epsilon$, as $\Delta \lambda$, indicating that anisotropy is driven primarily by the geometry of the contact network.  However we found that there is a big difference comparing pure/simple shearing with uniaxial compression.  The anisotropy parameters $\mu$ and $\Delta\lambda$ are smaller for uniaxial compression than for  shearing, by a factor of order 3 -- 5.  For pure/simple shearing, $\mu$ and $\Delta\lambda$ are monotonically decreasing as $\phi$ increases, while for uniaxial compression there is a kink with a sharp minimum at $\phi_J$.  

We have shown that $\Delta\lambda$ can be viewed as the nematic order parameter for bond directions in the contact network, while $\mu$ can be viewed as the nematic order parameter  of force weighted bonds in the the force network.  We have then generalized these to higher order, $2n$-fold orientational order parameters (tetratic, hexatic, etc.) for a more complete parameterization of the anisotropy of the configurations.  We find that, for $n>1$, these $2n$-fold orientational order parameters tend to be smaller for the force network as compared to the contact network, {\color{black}when compared relative to the $n=1$ nematic moment.  The adjustment of the forces on each bond tends to reduce higher order anisotropies.}

{\color{black}We then compared uniaxial compression to pure shearing, which both share the same geometric symmetry; the maximal and minimal stress axes for these two cases are in the same direction, and there is no other unique direction specified in the system.  We find that, while the nematic order parameters for these two cases are both quantitatively and qualitatively different (see Fig.~\ref{mu-lambda-vs-phi}), if we consider the higher order orientational moments measured relative to the nematic moment, then the two cases look quite similar (see Figs.~\ref{DP-vs-theta} -- \ref{An-A1-vs-phi}).  We thus conclude that, comparing uniaxial compression to pure shearing,  the main difference in system anisotropy is due to the nematic ordering.  

Finally, we compared pure shearing with simple shearing.  For our geometry, our simple shearing can be regarded as a superposition of pure shearing along the diagonal direction plus a system rotation.   In this case we found (see Fig.~\ref{mu-lambda-vs-phi}) that the nematic order parameters for these two cases are essentially equal.  However we found that the magnitude of the higher order orientational moments, while becoming equal  as $\phi$ increases towards $\phi_J^*$ and goes above, become increasingly different as $\phi$ decreases below $\phi_J^*$ (see Fig.~\ref{S2n-vs-phi-1234-ss}).   Moreover, the flow direction in simple shearing creates an additional special direction in the system, that can effect the orientations $\theta_{2n},\vartheta_{2n}$ of the order parameters.  While the lowest order  moments $n=1,2$ for simple shearing have orientations $\theta_2,\vartheta_2,\theta_4,\vartheta_4$  that are just rotated by $-\pi/4$ from those of pure shearing, the higher order moments seem to have no obvious relation between the two cases, even as one goes above jamming (see Fig.~\ref{theta-thetah-vs-phi-1234}).
It would be interesting to see if experiments on photoelastic disks \cite{Behringer} could detect the differences in the anisotropies of the contact and force networks, such as we find here.}

\begin{acknowledgments}
We thank Brendan Barrow for contributions at early stages of this work.
This work was supported by National Science Foundation Grant Nos. DMR-1809318 and PHY-1757062. Computations were carried out at the Center for Integrated Research Computing at the University of Rochester. 
\end{acknowledgments}

\section*{Appendix A: Compression Ensembles}
\label{appA}

In this appendix we describe in greater detail our compression ensemble and its limiting behaviors.  Our compressions start from random configurations of non-overlapping disks, constructed as described at the end of Sec.~\ref{MM}, at a given initial packing $\phi_\mathrm{init}$.  Here we will use smaller systems of $N=8192$ particles since the effect of varying $\phi_\mathrm{init}$ is greatest at smaller $\phi$, where finite size effects become negligible.  We will  focus on a single strain rate $\dot\epsilon=10^{-7}$, since at this rate one is in the linear rheology region ($p,\sigma\propto\dot\epsilon$) for $\phi\lesssim 0.82$, which covers the region of our primary interest.

\begin{figure}
\centering
\includegraphics[width=3.2in]{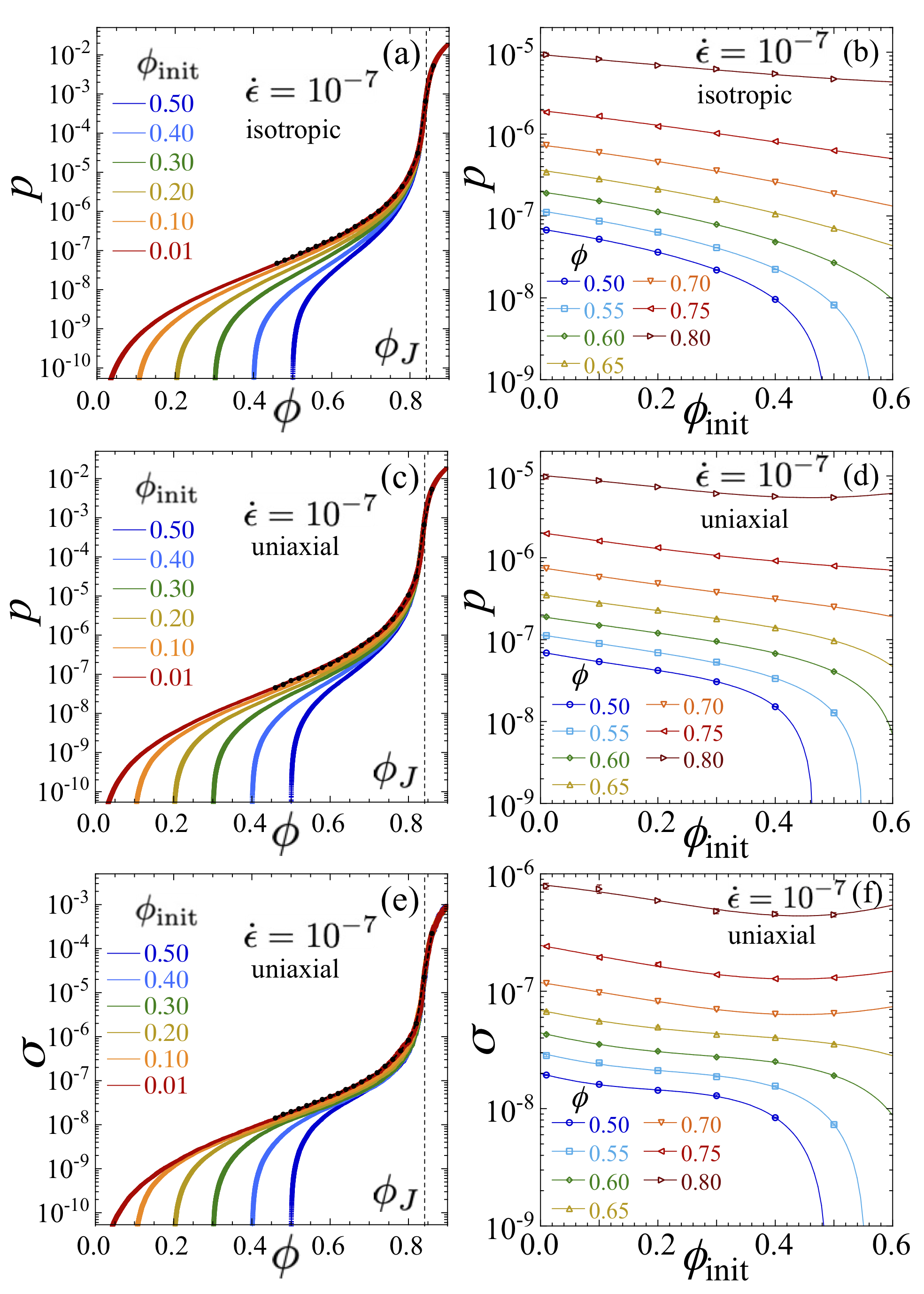}
\caption{(a) Pressure $p$ vs packing $\phi$ at a strain rate $\dot\epsilon=10^{-7}$, for isotropic compression of $N=8192$ particles starting from different initial packings $\phi_\mathrm{init}$.  The vertical dashed line indicates the compression-driving jamming $\phi_J = 0.8415$.  The width of each curve indicates the estimate error. (b) Data of panel (a) replotted as $p$ vs $\phi_\mathrm{init}$, at several different packings $\phi$.  The solid lines are cubic polynomial fits.  The black dots and dashed line in (a) are the extrapolated values of $p$ as $\phi_\mathrm{init}\to 0$, obtained from such fits.  (c) and (d) are analogous plots of $p$ for the case of uniaxial compression, while  (e) and (f) are the analogous plots of shear stress $\sigma$ for uniaxial compression.
}
\label{p-s-vs-phiinit}
\end{figure}

In Fig.~\ref{p-s-vs-phiinit}(a) we plot the resulting $p$ vs $\phi$ for isotropic compression, for values of $\phi_\mathrm{init}=0.01$ to 0.50.  In Fig.~\ref{p-s-vs-phiinit}(b) we replot these results as $p$ vs $\phi_\mathrm{init}$ at several different values of $\phi$.  Fitting the data of $p$ vs $\phi_\mathrm{init}$ to a cubic polynomial (shown as the solid curves in Fig.~\ref{p-s-vs-phiinit}(b)), we then extrapolate to determine the $\phi_\mathrm{init}\to 0$ limiting value of $p(\phi)$; these are shown as the black dots and dashed line in  Fig.~\ref{p-s-vs-phiinit}(a).  
The corresponding plots for uniaxial compression are shown in Figs.~\ref{p-s-vs-phiinit}(c) and \ref{p-s-vs-phiinit}(d) for the pressure $p$, and in Figs.~\ref{p-s-vs-phiinit}(e) and \ref{p-s-vs-phiinit}(f) for the shear stress $\sigma$.

We see that in all cases the stress (whether $p$ or $\sigma$) approaches a well defined limiting curve as $\phi_\mathrm{init}\to 0$.  There is a clear dependence of the stress on the particular value of $\phi_\mathrm{init}$ at small $\phi$, however this dependence goes away as $\phi$ increases, and the curves for all $\phi_\mathrm{init}$ approach the limiting curve as one enters the dense region just below jamming.  As $\phi$ decreases from this dense region, the curves for different $\phi_\mathrm{init}$ start to peel away from this liming curve, vanishing as $\phi\to\phi_\mathrm{init}$; the smaller is $\phi_\mathrm{init}$, the wider is the range of $\phi$ over which the finite $\phi_\mathrm{init}$ curve is  a good approximation for the $\phi_\mathrm{init}\to 0$ limiting curve.  


\begin{figure}
\centering
\includegraphics[width=3.2in]{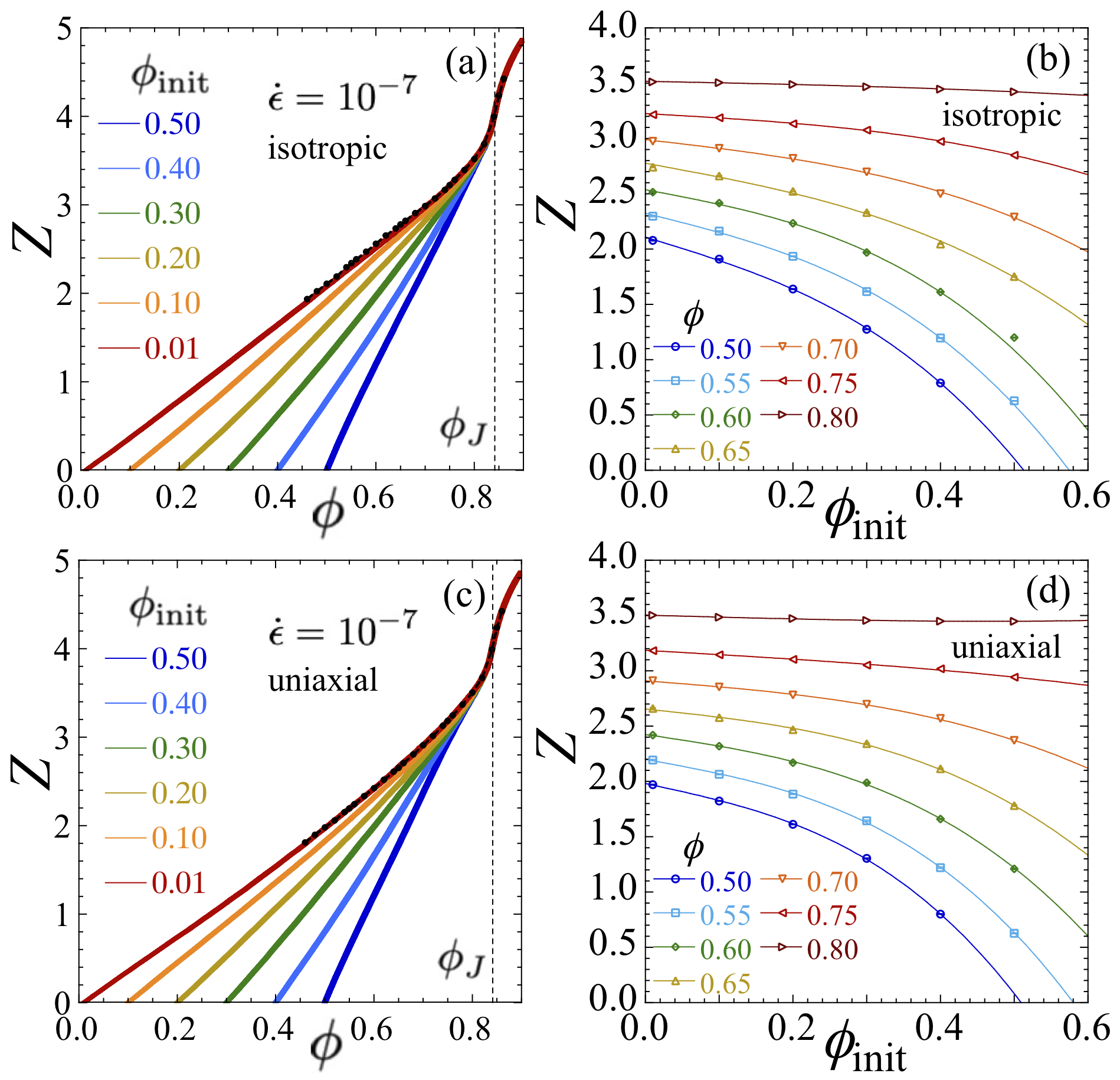}
\caption{(a) Average contact number $Z$ (including rattlers) vs packing $\phi$ at a strain rate $\dot\epsilon=10^{-7}$, for isotropic compression of $N=8192$ particles starting from different initial packings $\phi_\mathrm{init}$.  The vertical dashed line indicates the compression-driving jamming $\phi_J = 0.8415$.  The width of each curve indicates the estimate error. (b) Data of panel (a) replotted as $Z$ vs $\phi_\mathrm{init}$, at several different packings $\phi$.  The solid lines are cubic polynomial fits.  The black dots and dashed line in (a) are the extrapolated values of $Z$ as $\phi_\mathrm{init}\to 0$, obtained from such fits.  (c) and (d) are analogous plots of $Z$ for the case of uniaxial compression.
}
\label{Z-vs-phiinit}
\end{figure}

In Fig.~\ref{Z-vs-phiinit} we show similar plots of the average contact number per particle $Z$ for different $\phi_\mathrm{init}$.  We include rattler particles in our computation of $Z$ so that it remains well defined even at low $\phi$.  We see the same qualitative behavior as we found for the stress.  $Z$ approaches a well defined limit as $\phi_\mathrm{init}\to 0$, and the curves for finite $\phi_\mathrm{init}$ all approach this limiting curve as $\phi$ increases towards jamming.  From Figs.~\ref{p-s-vs-phiinit} and \ref{Z-vs-phiinit} we see that, for the $\phi_\mathrm{init}=0.40$ that we use in the main body of this work, effects due to the finite value of $\phi_\mathrm{init}$ should be rather small once $\phi\gtrsim 0.80$.
{\color{black}The behaviors shown in Figs.~\ref{p-s-vs-phiinit} and \ref{Z-vs-phiinit} suggest that $\phi_\mathrm{init}$ is an irrelevant variable in the sense of critical scaling, and that using a finite value for $\phi_\mathrm{init}$ will not effect the critical behavior at jamming, provided one restricts data to be sufficiently close to $\phi_J$.}

{\color{black}Such a conclusion is supported by earlier works by Ozawa et al. \cite{Ozawa} and Charbonneau and Morse \cite{CM}, who study inherent structures in systems of strictly hard-core spheres.  Starting from thermally equilibrated initial configurations at a packing $\phi_\mathrm{init}$, they carry out a {\em rapid} compression of the system to determine the value $\phi_J(\phi_\mathrm{init})$ at which the hard-core particles jam.  They find that, for initial packings below some threshold, $\phi_\mathrm{init}<\phi_\mathrm{th}$,  the jamming $\phi_J(\phi_\mathrm{init})$ is independent of $\phi_\mathrm{init}$ and  agrees with the random close packing value.    $\phi_J(\phi_\mathrm{init})$ starts to increase above this constant value only when $\phi_\mathrm{init}$ increases above $\phi_\mathrm{th}$.  The threshold $\phi_\mathrm{th}$ is associated with the glass transition found in mode coupling theory for the thermalized hard-core system; in two dimensions $\phi_\mathrm{th}\approx 0.7$.  All the $\phi_\mathrm{init}$ considered in our work are below this value.

The ensemble of thermalized equilibrium configurations from which \cite{Ozawa,CM} start their compressions is exactly the same as we use to start our compressions; for hard-core particles in thermal equilibrium at a fixed $\phi_\mathrm{init}$, all configurations in which there are no particle overlaps are equally likely.  However there are several differences between the models of \cite{Ozawa,CM} and our own, that might make one wonder how well their conclusions apply to our system.  They use hard-core particles, while we use soft-core particles.  Their system has a finite temperature $T$, while we are athermal with $T=0$.  They do a rapid compression, while we  compress at fixed rates $\dot\epsilon$, with jamming occurring in the quasistatic $\dot\epsilon\to 0$ limit. However, we will now argue that these two different approaches do indeed describe the same jamming critical point.

As shown in \cite{OTfiniteT}, for soft-core particles thermalized at a temperature $T$, and undergoing a strain deformation at a fixed rate $\dot\epsilon$, the dynamics of overdamped particles can be expressed in terms of the dimensionless parameters $k_e/T$ (the normalized particle stiffness) and $k_dV_sd_s^2\,\dot\epsilon /T$ (the P{\'e}clet number).  For the strictly hard-core particles considered in \cite{Ozawa,CM}, $k_e\to\infty$ and hence the only finite parameter is the P{\'e}clet number.  When these works compress rapidly, with the goal of avoiding thermalizing effects during compression, they are essentially doing simulations at large P{\'e}clet number, where the strain rate is much larger than the thermal relaxation rate.

In our model we are dealing with athermal soft-core  particles.  Here $k_e$ is finite but $T\to 0$.  Hence both the stiffness, $k_e/T$, and the P{\'e}clet number, $k_dV_sd_s^2\, \dot\epsilon/T$, diverge.  The ratio of these two, however, remains finite and gives the dimensionless strain rate, $k_dV_sd_s^2\,\dot\epsilon/k_e=\dot\epsilon\tau_0$.  The hard-core limit, where particle overlaps become negligible, is thus obtained by taking  $\dot\epsilon\to 0$.
Note, in the athermal soft-core model, the jamming critical point occurs at $(\phi_J, \dot\epsilon\to 0)$, and hence jamming is a property of this hard-core limit.
Thus, in both \cite{Ozawa,CM} and our present work,  jamming  is determined by the hard-core limit at large P{\'e}clet number.  Hence, the conclusions of \cite{Ozawa,CM}, that $\phi_J$ is independent of the $\phi_\mathrm{init}$ of the starting configurations, should imply that we get the correct critical jamming in our  athermal soft-core model, for any $\phi_\mathrm{init}$ that is not too big.
}

\begin{figure}
\centering
\includegraphics[width=3.2in]{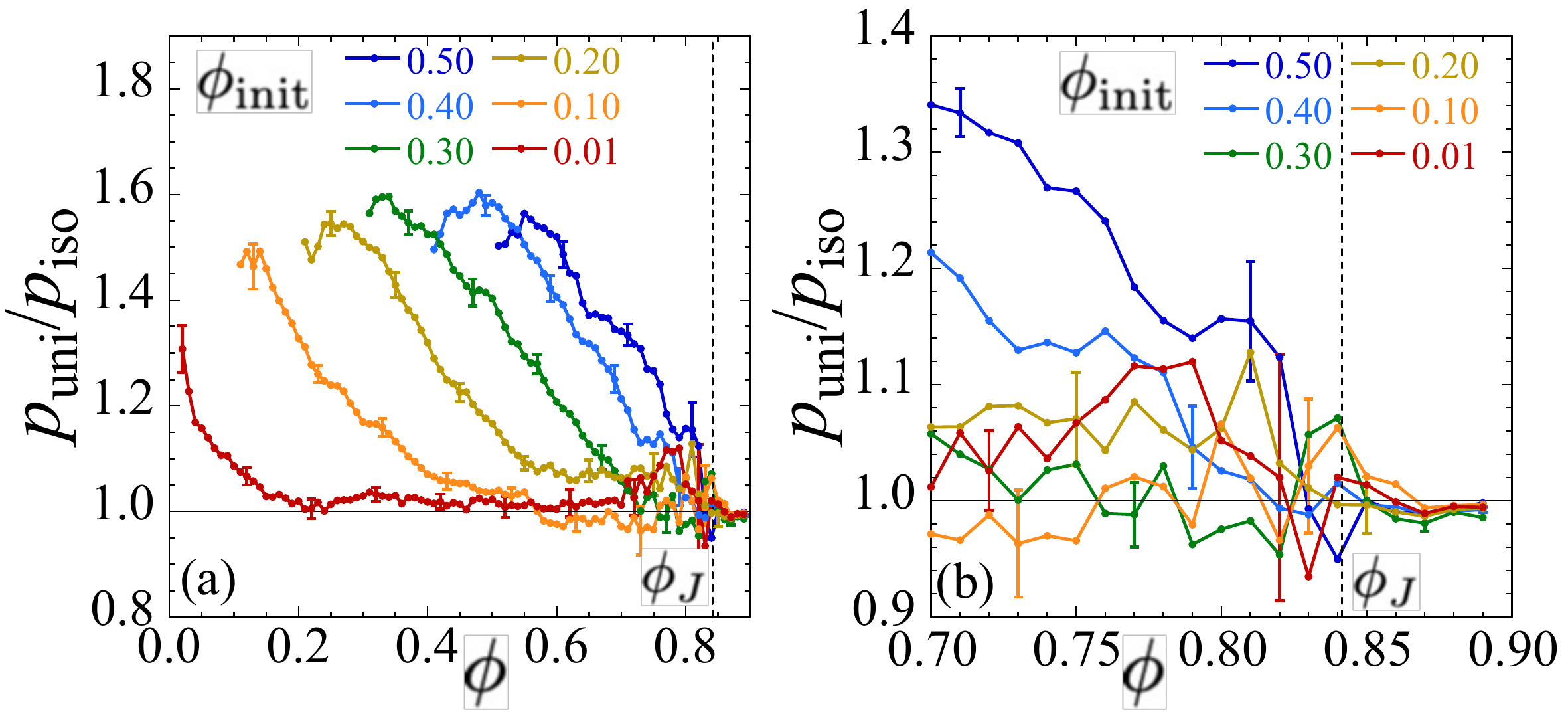}
\caption{Pressure ratio $p_\mathrm{uni}/p_\mathrm{iso}$ vs packing $\phi$, comparing uniaxial to isotropic compression, at a strain rate $\dot\epsilon=10^{-7}$ for $N=8192$ particles starting from different initial packings $\phi_\mathrm{init}$. (b) An expanded view of (a), looking closer in the vicinity of the jamming transition $\phi_J$.  The vertical dashed lines indicate the compression-driving jamming $\phi_J = 0.8415$.  For clarity, data points are shown only at intervals of $\Delta\phi=0.01$ and representative error bars are shown only on a subset of those points.
}
\label{Rp-vs-phi-phiinit}
\end{figure}

The results shown in Figs.~\ref{p-s-vs-phiinit} and \ref{Z-vs-phiinit} show qualitatively similar behavior, with respect to the dependence on $\phi_\mathrm{init}$, for both isotropic and uniaxial compression.  However we find an interesting result if we directly compare the pressure of the two cases.  In Fig.~\ref{Rp-vs-phi-phiinit} we plot the uniaxial to isotropic pressure  ratio $p_\mathrm{uni}/p_\mathrm{iso}$ vs $\phi$, for different values of $\phi_\mathrm{init}$.  Just below $\phi_J$ and above, we find $p_\mathrm{uni}/p_\mathrm{iso}=1$, within the estimated errors, as we reported  in Sec.~\ref{stress}.  However, as $\phi$ decreases, we see that $p_\mathrm{uni}/p_\mathrm{iso}$ eventually increases above unity.  This increase from unity shifts down to lower packings the smaller is the value of $\phi_\mathrm{init}$.  We conjecture that $p_\mathrm{uni}=p_\mathrm{iso}$ for all $\phi$, as $\phi_\mathrm{init}\to 0$.

\section*{Appendix B: Critical Scaling for Pure Shearing}
\label{appB}

In this appendix we provide more details of our pure shearing simulations, defined by the strain rate tensor of Eq.~(\ref{edef}).  Most prior work studying the effect of shearing on the jamming transition has considered simple shearing \cite{OT1,OT2,Vag,Heussinger1,Olsson3D,OT3,Vagberg.PRL.2014,Hatano2,Otsuki,Lerner,DeGiuli,Berthier,Andreotti}, 
where the strain rate tensor for flow in the $\mathbf{\hat x}$ direction is given by $\dot{\boldsymbol{\epsilon}}=\dot\epsilon \mathbf{\hat x}\otimes\mathbf{\hat y}$.  Simple shearing can be viewed as a superposition of pure shearing plus a system rotation with angular velocity $\dot\epsilon/2$.  Both simple and pure shearing preserve the system area.  

To pure shear, we  compress the system in the $\mathbf{\hat x}$ direction, while expanding it  in the $\mathbf{\hat y}$ direction, both at the same rate $\dot\epsilon/2$.  Unlike simple shear, where the system can be sheared indefinitely via the use of Lees-Edwards boundary conditions \cite{Lees}, we can only pure shear to a certain total strain $\epsilon=\dot\epsilon t$ before the system becomes too narrow in the $\mathbf{\hat x}$ direction and finite size effects become important.  For our system size of $N=32768$ particles, however, we find that we can always shear to at least  $\epsilon = 2$ with no apparent finite size effects, and that this is sufficient to reach  steady-state behavior.  

\begin{figure}
\centering
\includegraphics[width=3.4in]{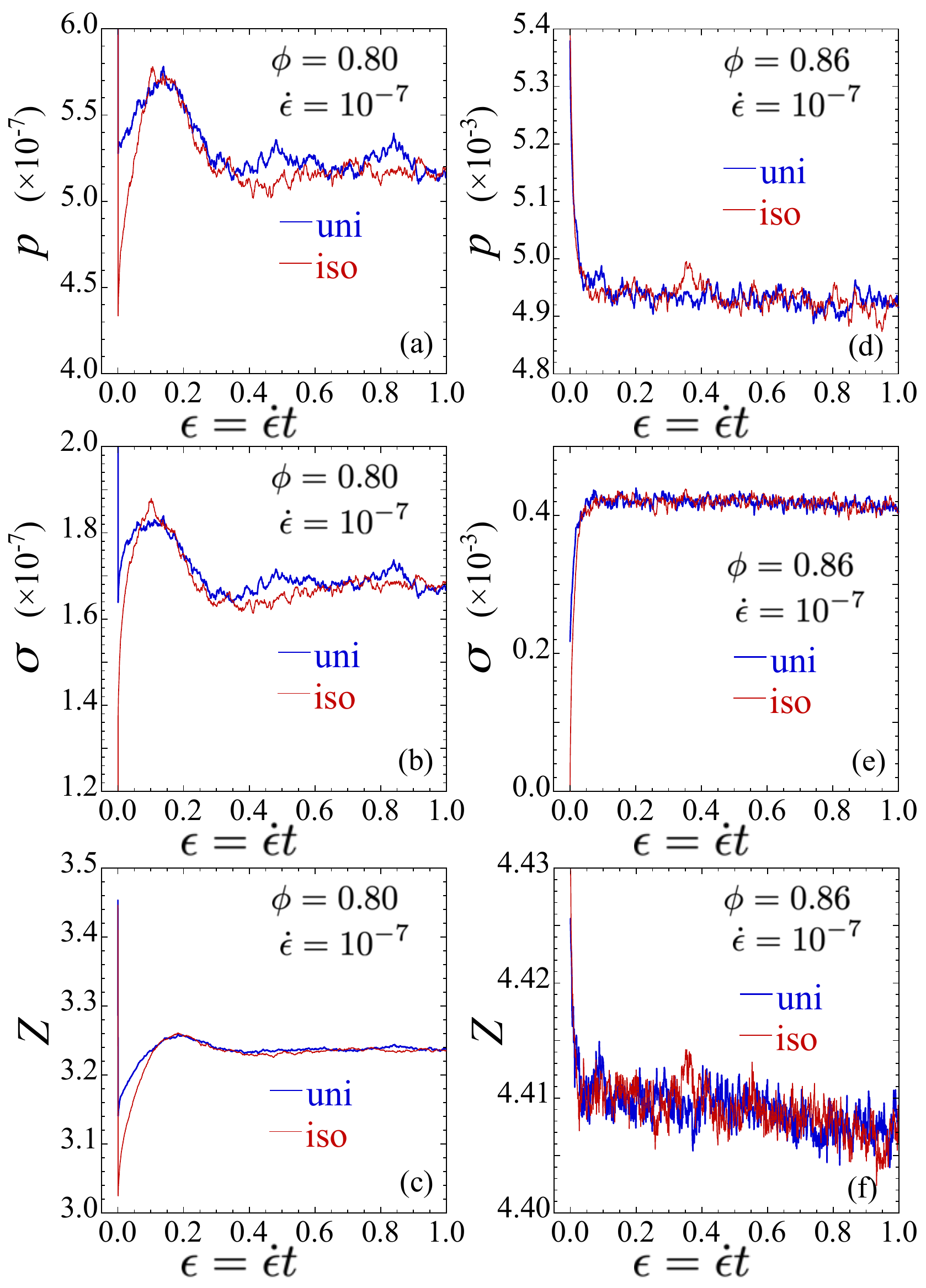}
\caption{(color online) (a) Pressure $p$, (b) shear stress $\sigma$, and (c) average contact number $Z$ (including rattlers) vs net strain $\epsilon=\dot\epsilon t$, as the system is pure sheared at $\phi=0.80 < \phi_J$ with the rate $\dot\epsilon=10^{-7}$.  Panels (d), (e), and (f) show similar results at $\phi=0.86 >\phi_J$.  Results are averaged over 10 independent initial configurations.  The thicker blue lines show results starting from initial configurations obtained from uniaxial compression, and so have a finite initial shear stress $\sigma>0$; the thinner red lines show results starting from initial configurations obtained from isotropic compression, and so have an initial $\sigma=0$. 
The system has $N=32768$ particles.
}
\label{p-s-z-vs-gamma}
\end{figure}

For the results reported in the main text, we pure sheared from an initial configuration  obtained from  uniaxial compression at the same rate $\dot\epsilon$.  For $\dot\epsilon >10^{-7}$ we averaged results over 10 independent  initial configurations, while for $\dot\epsilon \le 10^{-7}$ we used only a single initial configuration.  In all cases,  the reported steady-state values were obtained by averaging results over some strain interval $(\epsilon_1,\epsilon_2)$   within the steady-state region.  In contrast, to illustrate the evolution of the stress under pure shearing, in Fig.~\ref{p-s-z-vs-gamma} we show instantaneous results vs $\epsilon$ for configurations sheared  
at $\dot\epsilon=10^{-7}$, averaged over 10 independent initial configurations.  We compare the case where the initial configurations were obtained from uniaxial compression, and so have some finite initial shear stress $\sigma>0$, to the case where the initial configurations were obtained from isotropic compression, and so have $\sigma=0$.  

In Figs.~\ref{p-s-z-vs-gamma}(a), \ref{p-s-z-vs-gamma}(b), and \ref{p-s-z-vs-gamma}(c) respectively, we plot $p$, $\sigma$ and the contact number $Z$ (rattlers included) vs strain $\epsilon$ for $\dot\epsilon=10^{-7}$, at $\phi=0.80$ below jamming.  In Figs.~\ref{p-s-z-vs-gamma}(d), \ref{p-s-z-vs-gamma}(e), and \ref{p-s-z-vs-gamma}(f) we plot the same quantities at $\phi=0.86$ above jamming.
For the case of $\phi<\phi_J$, where the stress is due entirely to the finite strain rate, i.e., $p,\sigma\propto\dot\epsilon$, we find that the initial discontinuous change in the deformation (from uniaxial or isotropic compression to pure shear) results in an essentially instantaneous change in $p$, $\sigma$, and $Z$.  Following this initial instantaneous change, these parameters show a non-monotonic behavior as $\epsilon$ increases and  the system relaxes to its steady state.
We find this non-monotonic behavior to be limited to a fairly narrow window of $\phi$ below $\phi_J$.

For the case $\phi>\phi_J$, where there remains a finite stress even as $\dot\epsilon\to 0$, the initial change in $p$, $\sigma$ and $Z$ is still relatively rapid, though it is now smooth and continuous.  For  $\phi$ both above and below jamming, we see that, as $\epsilon$ increases, the system reaches a steady state, where these quantities plateau to roughly constant values.  The time needed to reach the steady state increases, and in principle diverges, as one approaches the jamming critical point, $\phi=\phi_J$ and $\dot\epsilon\to 0$.  We also see in Fig.~\ref{p-s-z-vs-gamma} that the values in the steady-state are independent of the starting initial configuration, as has been previously noted for simple shearing \cite{Vagberg.PRE.2011}.

\begin{figure}
\centering
\includegraphics[width=3.4in]{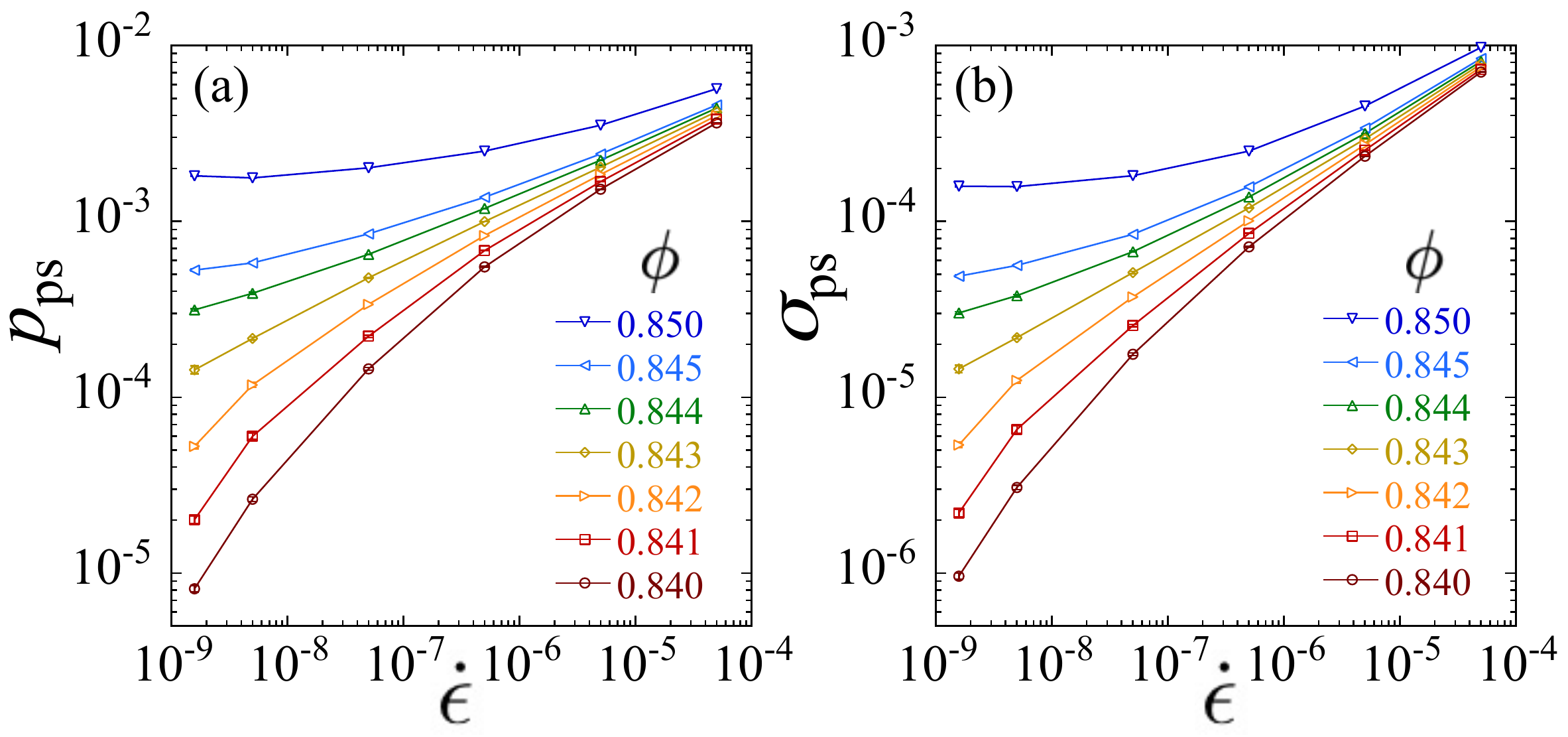}
\caption{(a) Pressure $p$ and (b) shear stress $\sigma$ vs strain rate $\dot\epsilon$, in steady-state pure shearing, for several different values of packing $\phi$ near jamming.  The system has $N=32768$ particles.
}
\label{p_ps-s_ps-vs-gdot}
\end{figure}

In Fig.~\ref{Rp-Rs-vs-phi} of Sec.~\ref{stress} we argued that the jamming packing fraction $\phi_J^*$ for pure shearing is slightly larger than the $\phi_J=0.8415$ for uniaxial or isotropic compression.  We now give further evidence for this.  A main characteristic of the jamming transition is that as $\dot\epsilon\to 0$, then below $\phi_J$ the stress $p,\sigma\to 0$ vanish, while above $\phi_J$ the stress $p,\sigma\to p_0,\sigma_0$ stays finite.  Thus, at small $\dot\epsilon$, curves of $p$ and $\sigma$ vs $\dot\epsilon$ will be concave for $\phi<\phi_J$, but convex for $\phi>\phi_J$.  In Fig.~\ref{p_ps-s_ps-vs-gdot} we plot the steady-state values of $p$ and $\sigma$ from pure shearing vs $\dot\epsilon$ for several different values of $\phi$ near jamming. Applying the above criterion to the pressure $p$ in Fig.~\ref{p_ps-s_ps-vs-gdot}(a), we clearly see that the jamming point for pure shearing satisfies $0.842<\phi_J^*<0.844$, and is thus larger than the jamming $\phi_J=0.8415$ found by us previously \cite{PeshkovTeitel} for isotropic compression.  The curves of shear stress $\sigma$ in Fig.~\ref{p_ps-s_ps-vs-gdot}(b) similarly argue for $0.842<\phi_J^*<0.844$, even though drawing conclusions from $\sigma$ can be complicated by larger corrections to scaling than exist for $p$ \cite{PeshkovTeitel2,OT2,VagbergOlssonTeitel}.

To determine the specific value of $\phi_J^*$ we can fit our data to the assumed critical scaling equation.  Since corrections-to-scaling have been found to be smaller for $p$ than for $\sigma$, we fit our data for pressure to the leading scaling form \cite{PeshkovTeitel2,OT2,VagbergOlssonTeitel},
\begin{equation}
p(\phi,\dot\epsilon)=\dot\epsilon^q f\left(\dfrac{\phi-\phi_J^*}{\dot\epsilon^{1/z\nu}}\right),
\label{escale}
\end{equation}
using the same fitting methods as detailed in Refs.~\cite{PeshkovTeitel2,OT2,VagbergOlssonTeitel}.  Plotting $p/\dot\epsilon^q$ vs $x\equiv(\phi-\phi_J^*)/\dot\epsilon^{1/z\nu}$ should then lead to a scaling collapse of the data to a common curve for different values of $\dot\epsilon$.

Since the scaling Eq.~(\ref{escale}) holds only asymptotically close to the jamming critical point $(\phi_J^*,\dot\epsilon\to 0)$, we wish to restrict the data used in the fit to  small values of $\dot\epsilon$ and values of $\phi$ near $\phi_J^*$.
We therefore use a data window similar to what we previously used \cite{PeshkovTeitel} for isotropic compression, with $\dot\epsilon \le 10^{-6}$ and $0.838\le\phi\le 0.848$.  Since our fitting procedure involves a polynomial expansion of the unknown scaling function $f(x)$, we also restrict the data used in the fit to $|x|\le 1$.

\begin{figure}
\centering
\includegraphics[width=3.4in]{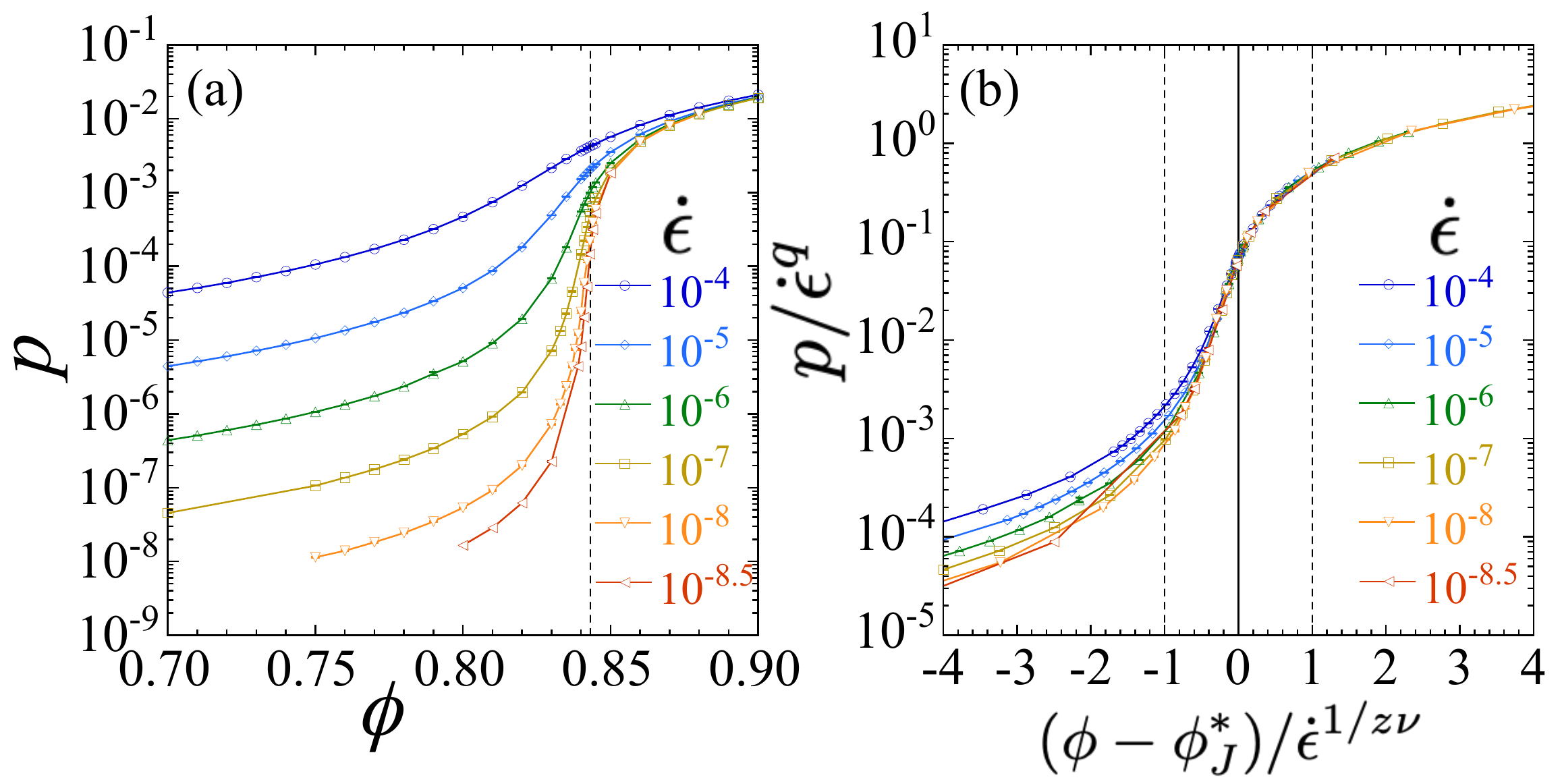}
\caption{(a) Pressure $p$ vs packing $\phi$ in steady-state pure shear, for different strain rates $\dot\epsilon$.  The vertical dashed line locates the jamming $\phi_J^*=0.84319$.  (b) Scaled pressure $p/\dot\epsilon^q$ vs scaled packing difference $(\phi-\phi_J^*)/\dot\epsilon^{1/z\nu}$, for different strain rates $\dot\epsilon$, according to the scaling Eq.~(\ref{escale}).  The scaling collapse is obtained using $\phi_J^*=0.84319$, $q=0.305$, and $1/z\nu=0.268$.  The system has $N=32768$ particles.
}
\label{p-vs-phi-scale}
\end{figure}

\begin{figure}
\centering
\includegraphics[width=3.4in]{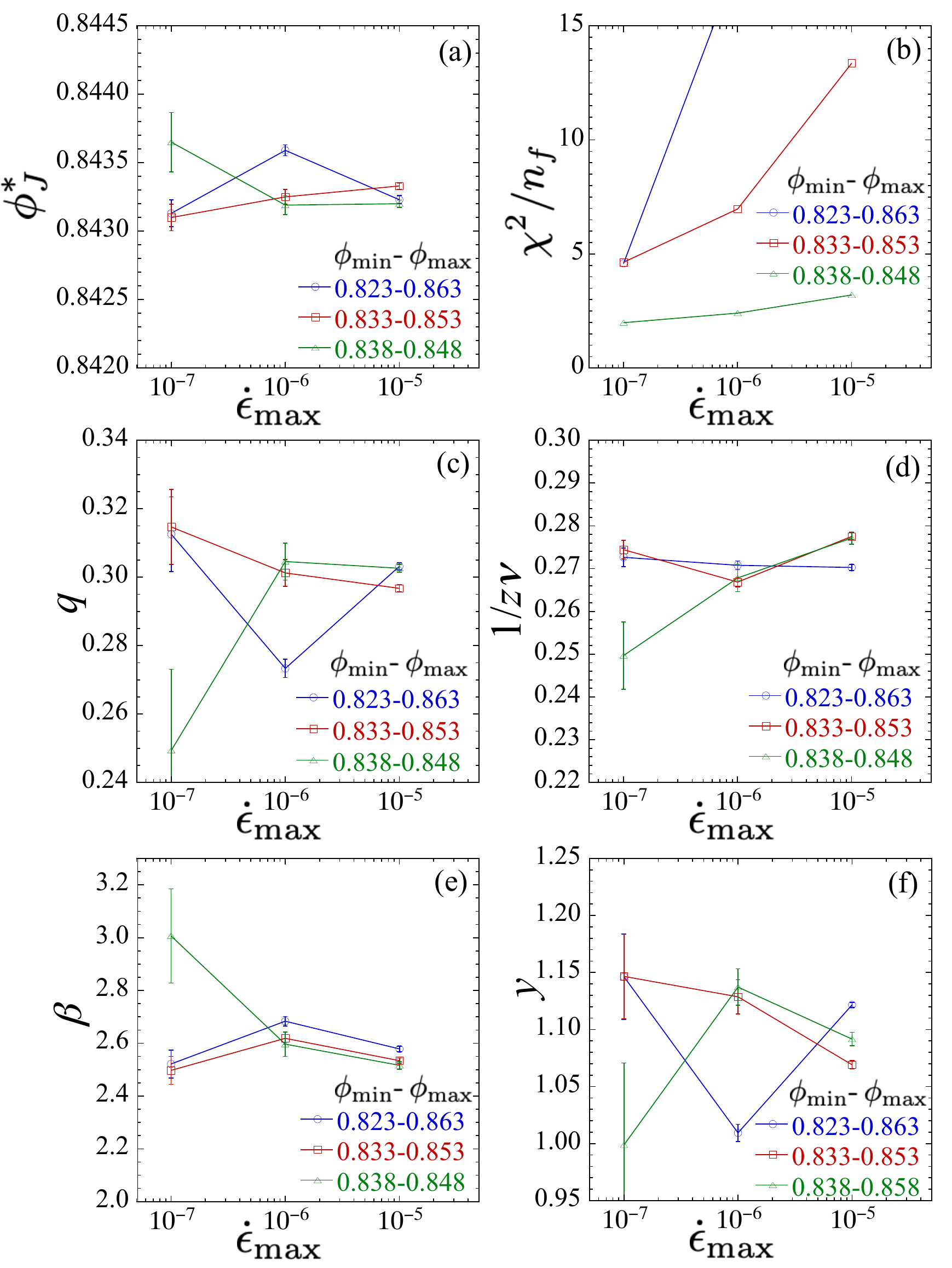}
\caption{Critical parameters obtained by fitting the data of Fig.~\ref{p-vs-phi-scale}(a) to the scaling form of Eq.~(\ref{escale}), for different windows of data.  Data windows are defined by $\dot\epsilon\le\dot\epsilon_\mathrm{max}$ and $\phi\in[\phi_\mathrm{min},\phi_\mathrm{max}]$. (a) The critical jamming packing $\phi_J^*$, (b) the $\chi^2$ per degree of freedom $n_f$ of the fit, (c) critical exponent $q$, (d) critical exponent $1/z\nu$, (e) bulk viscosity exponent $\beta=(1-q)z\nu$, and (f) yield pressure exponent $y=qz\nu$.
}
\label{fitparms}
\end{figure}

In Fig.~\ref{p-vs-phi-scale}(a) we plot our raw data $p$ vs $\phi$ for all our different values of $\dot\epsilon$.  In Fig.~\ref{p-vs-phi-scale}(b) we show the resulting scaling collapse, using the critical parameters obtained from our fit,  $q=0.305\pm 0.005$, $1/z\nu=0.268\pm0.003$, and  $\phi_J^*=0.8432\pm 0.0001$.  These result in \cite{PeshkovTeitel2}  the related exponent for the diverging bulk viscosity below jamming, $\lim_{\dot\epsilon\to 0}[p/\dot\epsilon]\sim (\phi_J^*-\phi)^{-\beta}$, with $\beta=(1-q)z\nu=2.60\pm0.05$, and the exponent for the vanishing pressure above jamming, $\lim_{\dot\epsilon\to 0}[p]\sim (\phi-\phi_J^*)^y$, with $y=qz\nu=1.14\pm 0.02$.
We find that the values of these parameters do not appreciably change if we slightly increase the window of data used for the fit, as shown in Fig.~\ref{fitparms}.  

We see that a good scaling collapse results from these parameters, that extends beyond the range of the data $|x|\le 1$ used to construct the fit.  Just as we found previously for isotropic compression \cite{PeshkovTeitel2,PeshkovTeitel}, the fit is excellent for $\phi>\phi_J^*$, but as $\phi$ decreases below $\phi_J$ we see  that the data splays away from the $\dot\epsilon\to 0$ limiting curve as $\dot\epsilon$ increases.  This is presumably due to corrections-to-scaling that become significant the further one moves from the jamming critical point \cite{PeshkovTeitel2,OT2,PeshkovTeitel}.

We can compare the above critical parameters with those found previously for isotropic compression and for simple shearing.  For isotropic compression in two dimensions we found previously \cite{PeshkovTeitel} $\beta=2.63\pm0.09$, $y=1.12\pm 0.04$, and $\phi_J=0.8415\pm 0.003$.  For simple shearing, the most accurate simulations \cite{OT2,OT3} (in our opinion)  give $\beta=2.77\pm 0.20$, $y=1.08\pm0.03$, and $\phi_J=0.84347\pm 0.00020$.  We thus find that the critical exponents $\beta$ and $y$ (and so also $q$ and $1/z\nu$) agree in all cases, within the estimated error; stress-isotropic jamming via isotropic compression has the same critical behavior as stress-anisotropic jamming via pure or simple shearing.  We also find that $\phi_J^*$ for pure shearing agrees with that found for simple shearing, and is slightly larger than the $\phi_J$ found for compression.


\clearpage
\bibliographystyle{apsrev4-2}

\end{document}